\documentclass{article}
\usepackage{graphicx}
\usepackage{amsmath} 
\usepackage{natbib}
\usepackage{url}
\usepackage{fullpage}
\title{Turbulent energy dissipation and intermittency in ambipolar diffusion magnetohydrodynamics}
\author{G.~Momferratos$^{1,2}\footnote{giorgos.momferratos@obspm.fr}$, P.~Lesaffre$^1$,
,E.~Falgarone$^1$, G.~Pineau~des~For\^ets$^{2,1}$\\
\\
$^1$ LRA/LERMA \'Ecole Normale Sup\'erieure/Observatoire de Paris,\\ 24, rue Lhomond
F 75231 Paris CEDEX 05\\
$^2$ IAS, Universit\'e de Paris-Sud, Centre universitaire d'Orsay
Bât 120 – 121 
91405 Orsay CEDEX}

\date{}

\begin{document}
\maketitle

\abstract{
  The dissipation of kinetic and magnetic energy in the interstellar
  medium (ISM) can proceed through viscous, Ohmic or ambipolar
  diffusion (AD). It occurs at very small scales compared to the
  scales at which energy is presumed to be injected. This localized
  heating may impact the ISM evolution but also its chemistry, thus
  providing observable features. Here, we perform 3D spectral
  simulations of decaying magnetohydrodynamic turbulence including the
  effects of AD. We find that the AD heating power spectrum peaks at
  scales in the inertial range, due to a strong alignment of the
  magnetic and current vectors in the dissipative range. AD affects
  much greater scales than the AD scale predicted by dimensional
  analysis. We find that energy dissipation is highly concentrated on
  thin sheets. Its probability density function follows a lognormal
  law with a power-law tail which hints at intermittency, a property
  which we quantify by use of structure function exponents. Finally,
  we extract structures of high dissipation, defined as connected sets
  of points where the total dissipation is most intense and we measure
  the scaling exponents of their geometric and dynamical
  characteristics: the inclusion of AD favours small sizes in the
  dissipative range.
}

\section{Introduction}


In partly ionized astrophysical fluids, magnetic fields remain
attached to the charged particles. The neutral fluid does not feel the
Lorentz force and usually drifts with respect to the charges. Only the
ion-neutral drag can remind the neutrals about the existence of
magnetic fields.  The fields are then able to slip through the bulk of
the fluid: this process is called ambipolar diffusion (AD).

\citet{mestel-spitzer-1956} were the first to realize its importance
in the context of star formation, where it would help magnetic fields
to diffuse out of a contracting dense core and allow it to form a
star.  \citet{Mullan-1971} discovered how ambipolar diffusion could
influence the dynamics of shocks which then yielded a wealth of papers
on the chemical signatures of C-type shocks, starting with
\citet{draine-et-al-1983} and \citet{FPH85}. \citet{Toth-1995}
produced the first multi-dimensional simulations with AD for the
stability of such shocks, and this opened the way to a collection of
analytical and numerical studies in various astrophysical
contexts. \citet{brandenburg-zweibel-1994} and
\citet{brandenburg-zweibel-1995} envisaged that ambipolar diffusion
could form very sharp structures, which would then induce the Ohmic
resistivity to reconnect the magnetic field in the interstellar medium
(ISM): they argued this could be a key element in solving
observational problems with the galactic dynamo
\citep{zweibel-brandenburg-1997}. In discs
\citet{blaes-balbus-1994,brandenburg-etal-1995} and
\citet{maclow-1995} showed AD was able to modulate the
magneto-rotational instability.  In the context of clouds and stars
formation
\citet{nakamura-li-2005,nakamura-li-2008,kudoh-basu-2008} and
\citet{kudoh-basu-2011} computed how the magnetic support of clouds
can leak out to let the gas condense and form dense cores and
stars. Finally, in the context of ISM turbulence
\citet{padoan-etal-2000,zweibel-2002,oishi-maclow-2006,lietalI,
  lietalII,lietalIII} and \citet{li-myers-mckee-2012} focused on how
magnetic fields decouple from the neutrals velocity or density and
estimated the heating resulting from the ion-neutral drift.

In the diffuse interstellar medium (ISM), turbulent energy dissipation
can be an important source of suprathermal energy driving hot
chemistry \citep{falgarone-puget-1995}. This may be evidenced by the
observed high values of the column density of species such as
$\mathrm{CH^+}$ and $\mathrm{SH^+}$. The formation of such species
requires energy barriers of the order of at least $2000~\mathrm{K}$ to
be overcome, in clouds where the average temperature is known to be a
few tens of $\mathrm{K}$. One possible explanation is that these cold
clouds contain pockets of hot gas heated by intermittent turbulent
dissipation. Hot chemistry is activated there, and it is possible to
construct models of turbulent dissipation that account for the high
column densities of $\mathrm{CH^+}$ \citep{godard-2008}.

In a turbulent magnetized fluid, dissipation can of course be due to
viscosity or resistivity, but when the fluid is partially ionized and
AD is at play, there can also be a significant contribution from the
heat released by ion-neutral friction, as demonstrated by several
authors in the context of the ISM
\citep{scalo-1977,zweibel-josafatsson-1983,elmegreen-1985,
  padoan-etal-2000,li-myers-mckee-2012}. Not only does the heating
help to raise the temperature which increases the rate of some
chemical reactions, but the ion-neutral drift velocity provides
additional energy in the reaction frame for ion-neutral reactions. In
some instances, this can open new chemical routes which would
otherwise be blocked by reaction barriers. The places of strong AD
heating are thus expected to bear specific chemical signatures such as
the ones encountered in magnetized vortices \citep{godard-2008} or
C-shocks \citep{lesaffre-et-al-2013}. We would hence like to
characterize the geometry and statistical properties of the regions of
strong turbulent dissipation in the ISM.

Before us, \citet{uritsky-et-al} conducted a thorough study of the
statistics of strong dissipation in the context of incompressible pure
magnetohydrodynamic (MHD) turbulence.  In this paper we make some
progress towards the physics of the ISM and we work with
incompressible MHD turbulence with or without AD. We stay within the
model of incompressible MHD in a first step to link our work with
\citet{uritsky-et-al} and to allow the use of spectral methods which
are well suited for the study of small-scale dissipative structures
because of their very low level of numerical dissipation.  In section
2 and 3 we briefly describe the equations and the numerical method
used and we present the simulations that were performed.  In section 4
we present an overall picture of the dissipation fields through the
time evolution of their average values, their pdfs and their
spectra. We also provide a qualitative view of the dissipation field
in physical space through color maps. Section 5 deals with the extreme
dissipative events, it begins with a discussion of the structure
functions of the velocity and magnetic fields and concludes with
results from the statistical analysis of the geometrical and dynamical
properties of structures of high dissipation. We discuss and conclude
our results in section 6.

\section{The equations}

\subsection{Ambipolar drift} 
In a partly ionized fluid, the time-dependent evolution of both the
neutral and the ionized fluids should in principle be
followed. However, in circumstances that we will make explicit below
(see subsection \ref{ADlengths}), we can adopt the strong coupling
approximation. In this approximation, we neglect the inertia, pressure
and viscosity of the ions in the ion momentum evolution and we are
left with the balance between the ion-neutral drag and the Lorentz
force

\begin{equation}
  \label{eq:ion-neutral-drift}
  \gamma \rho_i \rho_n (\mathbf{U}_i - \mathbf{U}_n) = \mathbf{J} \times \mathbf{B}
\end{equation}
where the current density
\begin{equation*}
  \mathbf{J}=  \frac{1}{4 \pi} (\nabla \times \mathbf{B}) \mbox{,}
\end{equation*}
$\rho_i$ and $\rho_n$ are the ion and neutral mass density
($\rho_i\ll\rho_n$), $\mathbf{U}_i$ and $\mathbf{U}_n$ are the ions
and neutrals respective velocities, and where $\gamma$ is the
coefficient of ion-neutral drag
\begin{equation*}
  \gamma = \frac{\langle \sigma v \rangle_{in}}{m_i + \mu}
\end{equation*}
with $m_i$ and $\mu$ the ions and neutrals mass per particle and
$\langle \sigma v \rangle_{in}$ the ion-neutrals collision
rate. Assuming that $n_{\mathrm{He}} = 0.2 n_{\mathrm{H_2}}$, we find
$\mu = 2.33 m_p$ for molecular gas, where $m_p$ is the mass of the
proton. In diffuse clouds, the average mass per ion is $m_i = 12 m_p$
as the dominant ion is $\mathrm{C}^+$. Following
\citet{draine-et-al-1983}, we take $\langle \sigma v \rangle_{in} =
1.9 \times 10^{-9} \mathrm{cm^3~s^{-1}}$ and we arrive at $\gamma=6.7
\times 10^{13} \mathrm{cm^{3}~s^{-1}~g^{-1}}$.

Within these approximations, the above balance
\eqref{eq:ion-neutral-drift} expresses the ion-neutral drift velocity
as a function of the magnetic field. When this is plugged into the
induction equation
\begin{equation*}
  \partial_t\mathbf{B} = \nabla \times (\mathbf{U}_i \times \mathbf{B}) + \eta \nabla^2 \mathbf{B}
\end{equation*}
one recovers the ambipolar diffusion term:
\begin{equation*}
  \nabla \times \left[ \frac1{\gamma \rho_i \rho_n} 
    (\mathbf{J} \times \mathbf{B}) \times \mathbf{B} \right]
\end{equation*}
\citep[cf.][]{balbus-terquem-2001}, which can be developed into
\begin{equation*}
  \nabla \times \left[ -\frac{B^2}{\gamma \rho_i \rho_n} 
    \mathbf{J} +\frac{\mathbf{J.B}}{\gamma \rho_i \rho_n} \mathbf{B} \right]
\end{equation*}
from which properly speaking only the first term takes the form of a
diffusion, with diffusion coefficient $\lambda_{\rm
  AD}=\frac{B^2}{\gamma \rho_i \rho_n}$
\citep{brandenburg-zweibel-1994}, but the qualificative is usually
retained for the whole term. In particular,
\citet{brandenburg-zweibel-1994} recognized that the second term
steepens the magnetic field profile near magnetic nulls. It should
finally be noted that AD itself is not per se able to reconnect the
field lines: this requires Ohmic diffusion.


\subsection{Incompressible MHD}

We now take $u_0$ a typical velocity and $l_0$ a typical length scale
as unit velocity and unit length to normalize our equations. We also
define $t_0=l_0/u_0$ as the unit of time.  We write the
non-dimensional velocity $\mathbf{u}=\mathbf{U_{\rm cdm}}/u_0$ where
$\mathbf{U_{\rm cdm}}$ is the center of mass velocity
\begin{equation*}
  \mathbf{U_{\rm cdm}} = \frac{\rho_i \mathbf{U}_i + \rho_n \mathbf{U}_n}
  {\rho}
\end{equation*}
with $\rho \sim \rho_n$ the total density of the gas, $\rho = \rho_i +
\rho_n$. We write the non-dimensional Alfv\'en velocity
$\mathbf{b}=\mathbf{B}/\sqrt{4 \pi \rho}/u_0$. The non-dimensional
current is simply $\mathbf{j}=\nabla \times \mathbf{b}$ where $\nabla$
is now understood as derivatives in coordinates in units of $l_0$:
$\nabla \rightarrow l_0\nabla$.


In the diffuse ISM, the sonic Mach number ${\cal M}_s = u_0 / c_s$
(where $c_s$ is the speed of sound) as well as the Alfv\'en Mach
number ${\cal M}_a = 1/|\mathbf{b}|$ take values in the range
$10^{-1}-10$ \citep{elmegreen-scalo-2004-1}. This wide range of values
suggests that although most of ISM turbulence is highly compressible,
incompressible turbulence is not irrelevant since in a weakly
compressible flow the density fluctuations $\Delta \rho / \rho \sim
{\cal M}_s^2$ are of the order of the square of the sonic Mach
number. Hence a turbulent flow with ${\cal M}_s<0.3$ can be adequately
described by the incompressible equations. For example, the studies of
\citet{brandenburg-zweibel-1995}, \citet{zweibel-2002} or
\citet{godard-2008} on turbulent dissipation with AD were all based on
the incompressible equations of motion.

We use the equations in \cite{balbus-terquem-2001} and the above
notations to derive the equations of incompressible, viscous,
resistive, AD MHD :
\begin{equation}
  \label{eq:mhd_adpaper}
  \begin{split}
    \partial_t \mathbf{u} &+ \left( \mathbf{u} \cdot \nabla \right)
    \mathbf{u} = -\nabla~p + \mathbf{j} \times \mathbf{b} + Re^{-1}
    \nabla^2 \mathbf{u}
    \\
    \partial_t \mathbf{b} &= \nabla \times \left ( \mathbf{u} \times
      \mathbf{b} \right) + Re_a^{-1} \nabla \times \big( \left(
      \mathbf{j} \times \mathbf{b} \right) \times \mathbf{b} \big) +
    Re_m^{-1} \nabla^2 \mathbf{b}
  \end{split}
\end{equation}
where $\mathbf{u}$ and $\mathbf{b}$ satisfy $\nabla \cdot \mathbf{u} =
0$ and $\nabla \cdot \mathbf{b} = 0$ and $p = P/ (u_0^2 \rho)$ is the
non-dimensional pressure, with $P$ the actual thermal
pressure. Equations \eqref{eq:mhd_adpaper} are parameterized by three
non-dimensional numbers $Re$, $Re_a$ and $Re_m$, for which we now give
estimates.

\subsection{Reynolds numbers}

Firstly, the kinetic Reynolds number $Re = u_0 l_0 / \nu$, where $\nu$
is the molecular viscosity of the fluid, expresses the relative
importance of inertial terms in comparison to the viscous term.  In
the neutral ISM, assuming that the most significant contribution to
viscosity is given by $\mathrm{H_2}$ collisions, we have $\nu \sim
\frac13 \lambda_{\mathrm{H_2}} c_s$ where $\lambda_{\mathrm{H_2}}$ is
the mean free path of the $\mathrm{H_2}$ molecule and $c_s = (\Gamma
k_B T/ \mu)^{1/2}$ is the isentropic sound speed, with $\Gamma \simeq
5/3$ the ratio of specific heats, and $k_B$ the Boltzmann constant.
The mean free path is given by $\lambda_{\mathrm{H_2}} \sim
(n_{\mathrm{H_2}} \sigma_{\mathrm{H_2}})^{-1}$ where
$n_{\mathrm{H_2}}$ is the number density of $\mathrm{H_2}$ and
$\sigma_{\mathrm{H_2}} = 3 \times 10^{-15}~\mathrm{cm}^2$ is an
estimate of the cross section of $\mathrm{H_2}$ collisions
\citep{monchick-schaefer-1980}. For molecular gas
$n_\mathrm{H_2}=0.5n_\mathrm{H}$ where $n_\mathrm{H}$ is the hydrogen
nuclei density. Under these assumptions, the kinetic Reynolds number
is of the order
\begin{equation*}
  Re \sim 1.8 \times 10^7 \left( \frac{n_{\mathrm{H}}}{100~\mathrm{cm^{-3}}} \right)
  \left( \frac{u_0}{1~\mathrm{km~s^{-1}}} \right) \left( \frac{l_0}{10~\mathrm{pc}} \right)
  \left( \frac{T}{100~\mathrm{K}} \right)^{-\frac{1}{2}} \mbox{.}
\end{equation*}
\cite{elmegreen-scalo-2004-1} quote typical values of the kinetic
Reynolds number in the cold ISM ranging from $10^5$ to $10^7$.

Secondly, $Re_m = u_0 l_0 / \eta$ is the magnetic Reynolds number,
where $\eta$ is the resistivity. The magnetic Reynolds number
expresses the relative importance of advection in comparison to Ohmic
diffusion in the dynamics of the magnetic field.  In a system with a
large value of the magnetic Reynolds number, the dynamics of the
magnetic field is dominated by advection and stretching.  The value of
the resistivity is given by
\begin{equation*}
  \eta = 234 \left( \frac{n}{n_e} \right) T^{1/2} \mathrm{cm^2 s^{-1}}
\end{equation*}
\citep{balbus-terquem-2001}, where $n$ is the total number density and
$n_e$ is the electron density. The order of magnitude of the magnetic
Reynolds number is
\begin{equation*}
  \begin{split}
    Re_m &= 2.2 \times 10^{17} \left( \frac{l_0}{10~\mathrm{pc}}
    \right) \left( \frac{u_0}{1~\mathrm{km~s^{-1}}} \right)  \times \\
    &\times \left( \frac{n_e}{10^{-4}n_\mathrm{H}} \right) \left(
      \frac{T}{100 ~\mathrm{K}} \right)^{-1/2}
  \end{split}
\end{equation*}
where we assumed $n=0.6n_\mathrm{H}$ for molecular gas.

Lastly, the AD Reynolds number $Re_a$ helps to measure the ratio of
the ambipolar to advective electromotive forces in the induction
equation:
\begin{equation}
  \label{eq:rad_adpaper}
  Re_a = \frac{t_0}{t_a},\qquad t_a=
  \frac{1}{\gamma \rho_i}
\end{equation}
where $t_a$ can be recognized in equation \eqref{eq:ion-neutral-drift}
as the ion-neutral friction time scale. The quantities $Re_a$ and
  $l_a$ should not be confused with their more usual definitions
  $R_{\rm AD}$ and $\ell_{\rm AD}$ as introduced by
  \cite{zweibel-brandenburg-1997}, for example. For instance, the
  usual values depend on the r.m.s. velocity and magnetic field,
  whereas our definition encompasses only the ion-drift time: see the
  subsection \ref{ADlengths} for more details. We can also write it as a
ratio of length scales
\begin{equation*}
  Re_a = \frac{l_0}{l_a}
\end{equation*}
where we define
\begin{equation}
  \label{eq:lad_adpaper}
  l_a=t_a u_0 
\end{equation}
which gives a typical length scale for ion-neutral decoupling.
The AD Reynolds number is an increasing function of the ionization
fraction $x = \rho_i/\rho$.  Using C$^+$ as the dominant ion of the
ISM, we find
\begin{equation*}
  Re_a = 4.9\times 10^3 \left( \frac{l_0}{10~\mathrm{pc}} \right)
  \left(  \frac{n_{\mathrm{C}^+}}{10^{-4}~n_\mathrm{H}} \right) \left( \frac{n_\mathrm{H}}{100~\mathrm{cm^{-3}}}  \right)
  \left(  \frac{u_0}{1~\mathrm{km~s^{-1}}} \right)^{-1}\mbox{.}
\end{equation*}



In the ISM, we have $Re_a \ll Re \ll Re_m $ which suggests the
ordering $l_a \gg l_\nu \gg l_\eta$ for the ambipolar, viscous and
resistive dissipation scales. However, our finite computing power does
not allow much dynamics of scales and here we can only afford $l_a >
l_D$ where $l_D = l_\nu \sim l_\eta$ is a single dissipative scale.
In this study the magnetic Prandtl number $\mathrm{Pr_m} = \nu / \eta$
is therefore taken equal to unity, so that the hydrodynamic and
magnetic Reynolds numbers are equal. This choice results in a single
dissipative range of scales for both the velocity and the magnetic
fields, a fact which simplifies the analysis considerably.


Although this numerical study is confined to Reynolds numbers that are
several orders of magnitude lower than those found in the interstellar
medium, it is relevant to ISM physics in the sense that it allows a
detailed quantitative study of the dissipation field as well as the
relative importance of the different types of turbulent dissipation,
which can all be important as a heating source for ISM chemistry.

\subsection{Lengths scales associated with AD}
\label{ADlengths}

We outline here various scales introduced by AD physics. The case of
C-shocks allows to clearly separate them.  We have already introduced
the length scale $l_a=u_0 t_a$ which corresponds to the re-coupling
length scale between ions and neutrals. It is the scale of variation
of the neutral's velocity in a C-shock with entrance velocity $u_0$,
which therefore has a length on the order of $l_a$ \citep{FP95}.

Under the strong coupling approximation, the scale of variation of the
{\it ions} velocity in the same C-shock is $l_{ai}=u_0 t_a /{\cal
  M}_a^2$ where ${\cal M}_a$ is the transverse Alfv\'enic Mach number
of the shock. For shocks with a large Alfv\'enic Mach number, this
length-scale is significantly smaller than $l_a$ and the structure of
the shock consists of a front in the ions velocity followed by a
smoother transition for the neutral velocity \citep[see Fig. 1 of ][
for example]{lietal06}.

In the case of typical ISM turbulence, though, ${\cal M}_a$ is of
order one, and both length scales do not differ significantly. In
fact, \citet{zweibel-brandenburg-1997} constructed the AD diffusion
Reynolds number of eddies of length scale $\ell$ and velocity $U$
based on the AD diffusion coefficient: $R_{\rm AD}(\ell)=\ell
U/\lambda_{\rm AD}=\ell/\ell_{\rm AD}$ where $\ell_{AD}=Ut_a/{\cal
  M}_a^2$ with ${\cal M}_a=U/c_a$.\footnote{For example, with
  $\ell=l_a$, we find that the AD Reynolds number of a C-shock is
  either $R_{\rm AD}=1$ or $R_{\rm AD}={\cal M}_a^2$ depending on
  whether we compute it with the post-shock or the pre-shock magnetic
  field strength.} \citet{zweibel-brandenburg-1997} then argue that
only eddies of length scales below $\ell_{\rm AD}$ should be affected
by AD.  We prefer to get a similar estimate by comparing the Fourier
amplitudes of the AD e.m.f.
($Re_a^{-1}(\mathbf{j}\times\mathbf{b})\times \mathbf{b} \rightarrow
Re_a^{-1} kb^3$) and the inertial e.m.f.  ($\mathbf{u}\times\mathbf{b}
\rightarrow ub$) in the the induction equation \eqref{eq:mhd_adpaper}. Wave
numbers above the critical wave number
\begin{equation}
  \label{eq:kad_adpaper}
  k_a=Re_a\sqrt{\langle u^2 \rangle}/ \langle b^2\rangle
\end{equation}
should be AD dominated. We hence define $\ell_a=2\pi/k_a$ accordingly,
as the length scale below which AD should be effective. Note that
$\ell_{\rm AD}$ and $l_0\ell_a$ differ only by a factor of $2\pi$.

Similarly, we can estimate the length scale below which the strong
coupling approximation breaks down by comparing the magnitude of the
neglected inertial term $D_t\rho_i\mathbf{U}_i$ to the coupling term
$\rho_n(\mathbf{U}_n-\mathbf{U}_i)/t_a$. Assuming that $\mathbf{U}_n$,
$\mathbf{U}_i$ and $\mathbf{U}_n-\mathbf{U}_i$ all share the typical
magnitude $u_0$, we then get a critical wavenumber $k_{\rm
  two-fluids}\simeq \rho_n/\rho_i/l_a$ above which the strong coupling
approximation fails and the two-fluids approximation is
needed. Provided $\rho_i/\rho_n$ is small (it is typically lower than
10$^{-3}$ if the main charges are C$^+$ ions), the strong coupling
approximation breaks down at scales much smaller than the typical AD
diffusion scale. Other authors
\citep{oishi-maclow-2006,padoan-etal-2000} have claimed that the
strong coupling approximation breaks down as soon as $\ell<\ell_{\rm
  AD}$ or $R_{\rm AD}<1$, where the ions inertia does not appear
explicitly. But Fig. 1 of \cite{lietal06} shows a C-shock computed
with the two-fluid approximation (solid) compared to an analytical
solution (dashed) using the strong coupling approximation, and the
agreement is perfect.  We hence believe that the strong coupling
approximation is a very good one in the low ionized ISM where
$\rho_i/\rho_n\ll 1$, even in cases where $R_{\rm AD}>1$. In
particular, $k_{\rm two-fluids}$ is at least a few ten times above the
largest wave number in all our AD simulations, which amply justifies
our use of the strong coupling approximation.

Finally, the observed emission of the ISM depends on the chemical and
thermal state of the gas, which are strongly linked to the heating.
The scale at which the heating takes place may not necessarily be
directly connected to the scale $\ell_a$ where AD undergoes a change
of dynamical regime. In fact we will see that it is not the case in
the present paper, and we will be forced to introduce yet an other
length scale $\ell^*_a$ for the typical thickness of sheets of strong
AD heating.

\section{The simulations}

\subsection{Method}
Many compressible methods have been devised to treat AD in the strong
coupling or in the two-fluids approximations: see \citet{Masson-2012}
and references therein.  Here, we solve the strong-coupling
incompressible equations \eqref{eq:mhd_adpaper} in 3D using a spectral method
for various values of the parameters and various initial
conditions. Our spectral code
ANK\footnote{\url{http://www.lra.ens.fr/~giorgos/ank}} is fully
de-aliased by use of the phase-shift method of
\citet{patterson-orszag-1971} and uses polyhedral truncation
\citep{canuto-et-al-1988}. Polyhedral truncation considers only
  these wave-vectors for which the sum of any two of their components
  does not exceed $2 N / 3$. Similarly with the widely used two-thirds
  rule, but contrary to an isotropic spherical truncation, this
  truncation scheme does not possess a sharp limit in wavenumber
  space. Polyhedral truncation allows us to keep 55 percent of the
modes active, in comparison to 33 percent for the standard two-thirds
rule, resulting in a more accurate description of the small
scales. The Fourier transforms are computed with FFTW with single
precision accuracy, and a standard fourth-order Runge-Kutta method is
used for time integration. We checked that our code gives the correct
solution for Alfv\'en waves damped by AD. As a resolution check
  for all simulations, we also looked for bumps in the kinetic and
  magnetic energy spectra near the truncation limit. In the case of
  the kinetic energy spectra we found bumps no larger than 15 \%,
  whereas no bumps were present in the magnetic energy spectra. Note
that we do not include a driving force in equations \eqref{eq:mhd_adpaper}:
our simulations are freely decaying. The MHD simulations with $512^3$
resolution take about $5000$ CPU hours until the peak of dissipation
but the equivalent AD-MHD simulations require ten times more CPU time
due to the more stringent time-step requirement.

In the spectral simulations performed, the velocity and magnetic
fields are defined on a regular Cartesian grid of points, while
boundary conditions are periodic in all directions.  Note the total
length of the computational domain is $2 \pi$ and the smallest
non-zero wave-vector has a norm of one.

\subsection{Initial conditions}

We use two types of initial conditions, corresponding to two different
situations for the magnetic and cross-helicities.

In the first case, the three lowest non-zero wave numbers of both the
velocity and the magnetic field are initially loaded with a
superposition of different Arnol'd-Beltrami-Childress \citep[ABC,
see][]{ABC} flows
\begin{equation}
  \label{eq:abc-flow}
  \begin{split}
    (u_x,u_y,u_z)&=(A \sin(k z) + C \cos(k y),B \sin(k x) \\
    &+ A \cos(k z), C \sin(k y) + B \cos(k x)) \mbox{.}
  \end{split}
\end{equation}
Different values of the coefficients $A,B,C$ are chosen for the first
three wave numbers from a uniform random number generator. In higher
wave numbers a random field with energy spectrum
\begin{equation}
  \label{eq:spectr-init}
  E(k) = C_E k^{-3} \exp \left(-2\left(k/k_c\right)^2\right),\quad k_c=3
\end{equation}
is superposed. The phases are chosen from a uniform random number
generator with the same seed for all simulations.

In the second case, the large scale initial flow is the Orszag-Tang
(OT) vortex
\begin{equation}
  \label{eq:orszag-tang-vortex}
  \begin{split}
    (u_x,u_y,u_z)&=(-2 \sin y,2 \sin x,0)\\
    (b_x,b_y,b_z)&=(-2 \sin(2 y) + \sin z,\\
    &2 \sin(x) + \sin z, \sin x + \sin y)
  \end{split}
\end{equation}
and in higher wave numbers a random velocity field with the same
properties as above is superposed. In order to keep the initial value
of magnetic helicity close to zero, no random magnetic field is added
to the OT initial condition, in contrast to the ABC initial condition.

The compressive components of the initial velocity and magnetic fields
are subtracted so that the initial condition is purely solenoidal.  In
all cases, the constant $C_E$ in equation \eqref{eq:spectr-init} is
chosen such that $\langle \mathbf{u}^2 \rangle= \langle \mathbf{b}^2
\rangle=1$, so that we start from equipartition between kinetic and
magnetic energy.  The energy of the initial condition fields is
concentrated on large scales $k < k_c$ due to the exponential cutoff
in \eqref{eq:spectr-init}.

In the case of the ABC initial condition, the non-dimensional
cross-helicity
\begin{equation*}
  H_c = \frac{2 \langle \mathbf{u} \cdot \mathbf{b} \rangle}{
    \sqrt{\langle \mathbf{u}^2 \rangle 
      \langle \mathbf{b}^2 \rangle} } 
\end{equation*}
is $\sim 2 \times 10^{-3}$, corresponding to a low initial correlation
between the velocity field and the magnetic field. The mean magnetic
helicity
\begin{equation*}
  H_m = \langle \mathbf{a} \cdot \mathbf{b} \rangle
\end{equation*}
where $\mathbf{a}$ is the vector potential with $\mathbf{b} = \nabla
\times \mathbf{a}$, is considerable, $\sim 0.2$. In the case of the OT
initial condition the non-dimensional cross-helicity is $\sim 0.1$
while the mean magnetic helicity is almost zero, $\sim 1 \times
10^{-9}$. Thus these two different initial conditions represent
evolution under different constraints: in the ABC case, low
cross-helicity and sizable magnetic helicity whereas in the OT case
sizable cross-helicity but low magnetic helicity. This fact is
important because in the ideal MHD limit (inviscid and non-resistive)
the energy, cross-helicity and magnetic helicity are all conserved
during the evolution. If AD is included in the ideal MHD equations,
the conservation of magnetic helicity remains while energy and
cross-helicity conservation are broken. This is a consequence of the
form of the AD term in the induction equation, which takes the form of
an advection term
\begin{equation*}
  \nabla \times \left( \mathbf{u}_d \times \mathbf{b} \right),\qquad
  \mbox{with}~~\mathbf{u}_d = Re_a^{-1} (\mathbf{j} \times \mathbf{b})
\end{equation*}
the non-dimensional ion-neutral drift velocity.  This form also
implies that although AD is a dissipative process, it conserves
magnetic flux and is thus unable to reconnect field lines.

\begin{table}
  \begin{center}
      \begin{tabular}{|c || c | c | c | c | c |c | c | c|}
        \hline
        \# & $N$ & $L$ &$\lambda$ & $l_d$ & $Re_\lambda$ & $Re = Re_m$ 
        & $Re_a$ & Initial condition\\
        \hline
        1 &  128 & 2.60 &  1.30 & 0.0607 & 189 &  219 &   - & ABC \\
        \hline
        2 &  128 & 2.59 &  1.31 & 0.0605 & 210 &  219 &   - &  OT \\
        \hline
        3 &  128 & 2.70 &  1.44 & 0.0390 & 217 &  219 & 100 & ABC \\
        \hline
        4 &  128 & 2.72 &  1.44 & 0.0381 & 228 &  219 & 100 &  OT \\
        \hline
        5 &  256 & 2.46 &  0.94 & 0.0375 & 353 &  551 &   - & ABC \\
        \hline
        6 &  256 & 2.46 &  0.91 & 0.0389 & 377 &  551 &   - &  OT \\
        \hline
        7 &  256 & 2.62 &  1.08 & 0.0185 & 412 &  551 & 100 & ABC \\
        \hline
        8 &  256 & 2.66 &  1.09 & 0.0196 & 444 &  551 & 100 &  OT \\
        \hline
        9 &  512 & 2.28 &  0.67 & 0.0237 & 591 & 1374 &   - & ABC \\
        \hline
        10 &  512 & 2.12 &  0.58 & 0.0245 & 604 & 1374 &   - &  OT \\
        \hline
        11 &  512 & 2.49 &  0.83 & 0.0091 & 756 & 1374 & 100 & ABC \\
        \hline
        12 &  512 & 2.36 &  0.75 & 0.0095 & 750 & 1374 & 100 &  OT \\
        \hline
        13 &  512 & 1.84 &  0.58 & 0.0074 & 640 & 1374 &  10 & ABC \\
        \hline
        14 &  512 & 2.80 &  1.00 & 0.0084 & 927 & 1374 &  10 &  OT \\
        \hline
      \end{tabular}
    \caption{Parameters of the simulations. $N$: linear resolution,
      $L$: integral length scale at the peak of dissipation (pd) ,
      $\lambda$: Taylor microscale (pd), $l_d$: dissipative scale
      (pd), $Re_\lambda$: Taylor microscale Reynolds number $U \lambda
      Re$ (pd), $Re$: kinetic Reynolds number, $Re_m$: magnetic
      Reynolds number, $Re_a$: AD Reynolds number.\label{tab:sims}
      }
  \end{center}
\end{table}

\subsection{Parameters}

The parameters of the simulations performed are shown in table
\ref{tab:sims}. Throughout this paper, we focus mainly on the analysis
of the OT initial condition with $Re_a = 10,100,\infty$, and we
discuss the differences with respect to the ABC initial conditions
only when they arise. The Taylor microscale Reynolds number
$Re_\lambda$ is defined as $Re_\lambda = U \lambda Re$ where $ U
=\sqrt{ \langle \mathbf{u}^2 \rangle}$ is the r.m.s. velocity and
\begin{equation*}
  \lambda = 2 \pi \left( \frac{\int_0^\infty e(k)~dk}
    {\int_0^\infty k^2 e(k)~dk} \right)^{\frac{1}{2}}
\end{equation*}
is the Taylor microscale, with $\int_0^\infty e(k)~dk=\frac12 \langle
\mathbf{u}^2 +\mathbf{b}^2 \rangle $ the total energy and $e(k)$ the
total energy spectrum. The value of $Re_\lambda$ is given at the peak
of viscous plus Ohmic dissipation where

\begin{equation*}
  \begin{split}
    \langle \varepsilon \rangle &= \langle \varepsilon_o \rangle +
    \langle \varepsilon_v \rangle 
    \\
    \varepsilon_o &= Re_m^{-1} \mathbf{j}^2
    \\
    \varepsilon_v & = \frac{Re^{-1}}{2}
    \sum_{i,j=1}^{3}\left(\partial_i u_j + \partial_j u_i \right)^2
    \mbox{.}
    \\
  \end{split}
\end{equation*}
The time when the peak of dissipation occurs is appropriate for
analysis because for a given integral length scale
\begin{equation}
  \label{eq:ils}
  L = 2 \pi \frac{\int_0^\infty k^{-1} e(k)~dk}{\int_0^\infty e(k)~dk}
\end{equation}
and dissipative scale (assuming Kolmogorov scaling)
\begin{equation}
  \label{eq:etad}
  l_d = \left(\frac{Re^{-3}}{\langle \varepsilon 
      \rangle}\right)^{\frac{1}{4}} 
\end{equation}
the scale separation $L/l_d$ between the energy-containing scales and
the dissipative scales is maximum. Another desirable property at the
peak of dissipation is quasi-stationarity, due to the time derivative
of the dissipation rate which cancels at the peak, by definition. In
all the MHD simulations (with $Re_a^{-1}=0$), two snapshots of the
fields were recorded for analysis: one at the peak of dissipation and
a second one one eddy turnover time later.  We define the macroscopic
eddy turnover time as
\begin{equation}
  \label{eq:ett}
  T = \sqrt{3}\frac{L}{U}
\end{equation}
where the one-directional r.m.s. velocity $U/\sqrt{3}$ and the
integral length scale $L$ were both computed at dissipation peak to
estimate when to output the next snapshot.  For the AD simulations
(with $Re_a^{-1}>0$), we could not afford to compute beyond the
dissipation peak.

\subsection{Power-spectra}
\label{ranges}

\begin{figure}
  \centering
  \begin{minipage}[h]{0.5\linewidth}
    \includegraphics[width=\textwidth]{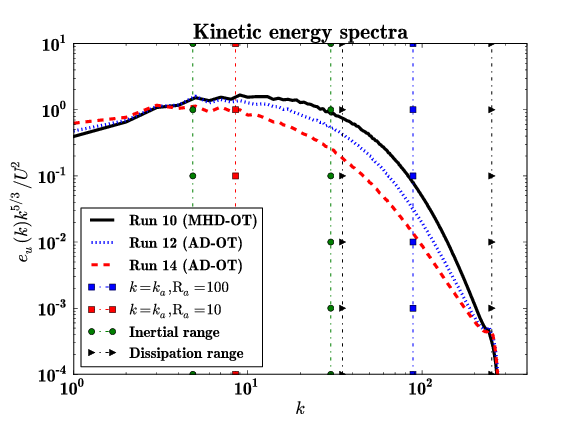}
  \end{minipage}
  
  \begin{minipage}[h]{0.5\linewidth}
    \includegraphics[width=\textwidth]{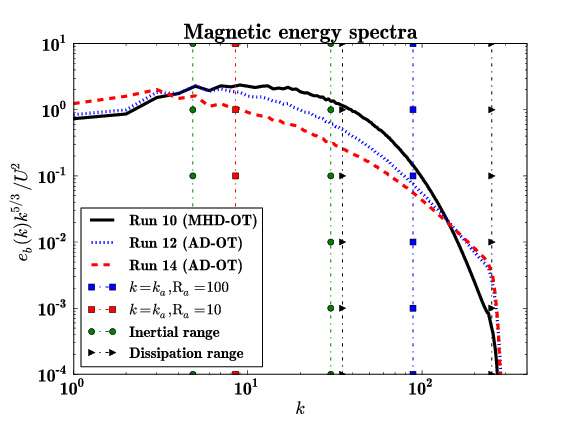}
  \end{minipage}
\caption{  Compensated kinetic (top) and magnetic (bottom) energy
    spectra for OT runs 10, 12 and 14 at the temporal peak of 
    dissipation. The assumed limits of the inertial and dissipation
    ranges are shown in dashed vertical lines. The vertical blue and red
    dashed-dotted lines with square symbols correspond to the expected AD critical wavenumber $k_a$ (see equation \eqref{eq:kad_adpaper}) in runs 12 and 14, respectively.
    \label{fig:spectra}}
\end{figure}

The kinetic and magnetic energy spectra of the high resolution OT runs
10-12-14 are shown in Figure \ref{fig:spectra}, at the temporal peak
of dissipation. The spectra are compensated by the Kolmogorov law
$k^{-5/3}$ and normalized by $U^2$.  The extent of the inertial range,
as defined by the portion of the spectra that has slope $-5/3$ is very
limited, especially in the cases with AD.  In the same Figure we show
the limits of the inertial and dissipation ranges assumed for the
analysis of section \ref{sect:extraction}: they are taken from
\cite{uritsky-et-al} as $[0.21,1.3]$ and $[0.025,0.18]$ respectively
(in units of $l_0$).

The kinetic and the magnetic energy appear to remain in approximate
equipartition across all scales except for the smallest scales in the
AD runs where magnetic energy dominates the kinetic energy (the
tick-marks on the vertical lines can guide the eye to estimate the
relative position of the curves between the upper and the bottom
panel).

The vertical dash-dotted lines with square symbols correspond to the
wave-number $k_a$ defined in equation \eqref{eq:kad_adpaper}, for the cases of
runs 12 and 14.  Surprisingly, departures from MHD spectra start at
about the same wave number for all AD runs (in the range $k \sim 5-8$
for both the kinetic energy spectrum and the magnetic one): this hints
at the fact that $\ell_a=2\pi/k_a$ is not the proper scale to assess
the dynamical importance of AD in our simulations. In particular,
dynamics can be affected at scales much larger than that in run 12.
Although the difference between the pure MHD and AD MHD spectra is
modest, there is a clear tendency for AD to flatten the energy
spectra, especially for the magnetic energy.  This is in line with the
idea by \citet{brandenburg-zweibel-1994} that AD diffuses magnetic
fields on the one hand, but on the other hand helps to build sharper
magnetic structures in specific places.



\section{The dissipation field}

\subsection{Total dissipation}

\begin{figure}
  \centering
  \includegraphics[width=0.5\textwidth]{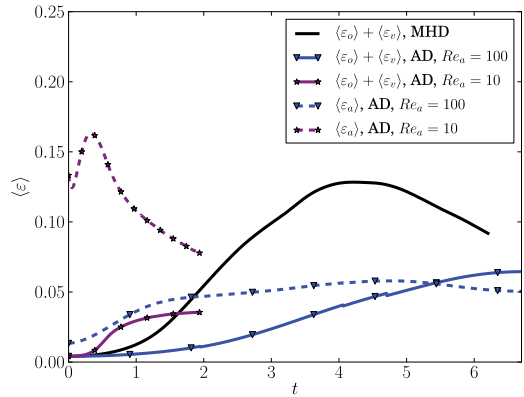}
  \caption{Time evolution of volume integrated dissipation rates for
    the OT runs 10, 12 and 14. Solid lines show the Ohmic plus viscous
    dissipation which we use to define peak dissipation.
  \label{fig:diss}
}
\end{figure}

We present in Figure \ref{fig:diss} the time evolution of volume
integrated dissipation rates. The viscous and Ohmic dissipation rates
follow each other closely and we don't separate their respective
contributions in this figure.  Most of the time, the total dissipation
rates due to viscosity, resistivity and AD are of comparable
magnitude.  But the total AD dissipation rate is seen to peak before
the Ohmic plus viscous dissipation rate, especially at low values of
$Re_a$. This makes our choice of the temporal peak of Ohmic plus
viscous dissipation more appropriate to avoid the initial transient
spike of AD dissipation (this spike is even more pronounced in the ABC
case run 13).

We note that Ohmic dissipation is not enhanced by the presence of
AD. On the contrary, the peak value of the Ohmic plus viscous
dissipation decreases as $Re_a$ is decreased. Even though
\citet{brandenburg-zweibel-1994}'s idea that AD sharpens magnetic
structures at small scales is valid in our simulations, AD also
smooths the fields at intermediate scales and the net effect on the
global Ohmic heating is to decrease it. However, this may be due to
the finite dynamical range in our simulations. Higher $Re_m$ Reynolds
number simulations, if they yield enhanced magnetic power in a more
extended range at small scales, could result in a globally enhanced
rate of reconnection, in agreement with
\citet{zweibel-brandenburg-1997}'s model.

\subsection{Probability distribution function \label{sect:pdf}}

\begin{figure}
  \centering
  \includegraphics[width=0.5\textwidth]{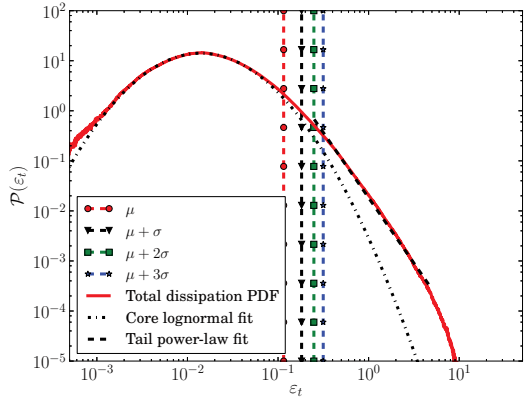}
  \caption{ Log-normal core and power-law tail fit for the pdf of the
    total dissipation (Run 12 AD - OT). The core follows a
    log-normal distribution with mean $\mu_l \simeq -4.27$ and standard
    deviation $\sigma_l \simeq 1.03$ while the tail follows a
    power-law with exponent $-\tau \simeq -2.61$. Vertical lines show
    the mean value (red) and thresholds located at 1 (black), 2 (green)
    and 3 (blue) standard deviations above the mean value.
  \label{fig:pdf}
}
\end{figure}

The probability density function (pdf) of the total dissipation rate
\begin{equation*}
  \varepsilon_t    =  \varepsilon_o  + \varepsilon_v + \varepsilon_a 
\end{equation*}
where
\begin{equation*}
  \varepsilon_a  = Re_a^{-1}(\mathbf{j} \times \mathbf{b})^2
\end{equation*}
is shown in Figure \ref{fig:pdf} for the high-resolution run 12.  The
core of the pdf is very close to the log-normal distribution
\begin{equation*}
  \mathcal{P}_c(\varepsilon_t) \propto \exp \left( - \frac{(\ln \varepsilon_t - 
      \mu_l)^2}{\sigma_l^2} \right)
\end{equation*}
with mean $\mu_l \simeq -4.27$ and standard deviation $\sigma_l \simeq
1.03$, while the tail of the distribution can be fitted by a power-law
\begin{equation*}
  \mathcal{P}_t(\varepsilon_t) \propto \varepsilon_t^{-\tau}
\end{equation*}
with exponent $\tau \simeq 2.61$. This power-law is one of the
signatures of intermittency of dissipation \citep{frisch-1995}.  For
still higher values of the total dissipation the pdf has an
exponential cut-off, although these high dissipation values are
close to the sampling limit. In the analysis of the next
section (extraction of structures of high dissipation) effectively
only the power-law range of the distribution is sampled.

\begin{figure}
  \centering
  \includegraphics[width=0.5\textwidth]{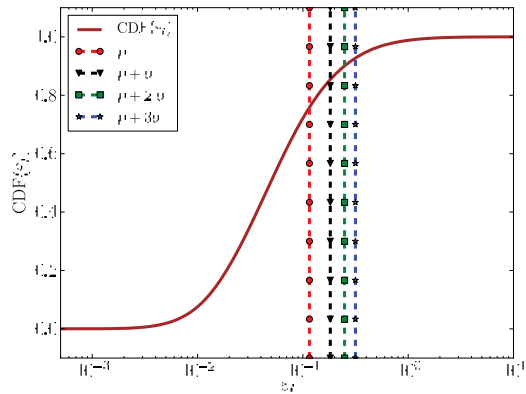}
  \caption{Cumulative probability density function of the total
    dissipation for run 12 (AD - OT).\label{fig:cdf}}
\end{figure}

In Figure \ref{fig:cdf} we present the cumulative probability density
function of the total dissipation for run 12. It flattens out before
reaching unity, which shows that high values of the dissipation are
concentrated in a small volume subset of the spatial domain. The
analysis of the next section concerns events that take place in this
high-dissipation plateau.

\subsection{Power-spectrum}

\begin{figure}
  \centering
  \includegraphics[width=0.5\textwidth]{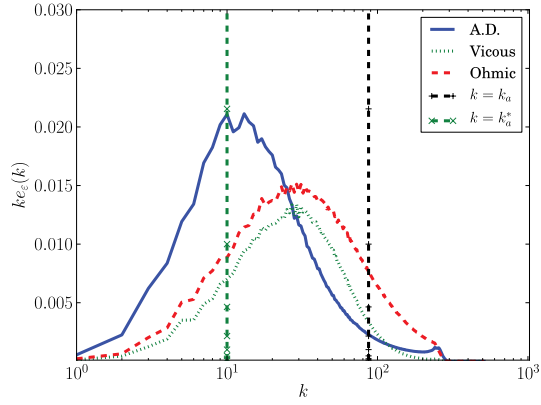}
  \caption{Compensated dissipation spectra for run 12 (AD-OT, $Re_a =
    100$). Blue solid line: AD dissipation, red dashed line: Ohmic
    dissipation, green dotted line: viscous dissipation. We plot $k
    e_\epsilon(k)$ in a log-lin plot, so that the area under the curve
    over any interval shows directly the amount of power inside this
    interval.  We mark the position of the maximum value of $k
    e_{\epsilon_a}$, at $k=k_a^*=2\pi/\ell_a^*$, and the position of
    $k=k_a$.\label{fig:diss-spectra}}
\end{figure}

We now investigate the distribution of the energy dissipation with
respect to spatial scale. This can be discerned thanks to the power
spectra of the dissipation fields, displayed in figure
\ref{fig:diss-spectra}.  The AD heating peaks at larger scales in
comparison to Ohmic and viscous dissipation. The scale of this peak
$\ell_a^*$ is actually only a few times smaller than the integral
length scale. At this range of scales AD dissipation is much more
important than Ohmic and viscous dissipation. This suggests that the
heating due to AD has a characteristic length scale $\ell_a^*$, which
can be much larger than the dimensional estimate $\ell_a=2\pi/k_a$
\eqref{eq:kad_adpaper} for run 11. This length scale is relevant to the
heating, and hence may manifest itself in structures revealed by
chemical tracers of the warm chemistry of the ISM.

\begin{figure}
  \centering
  \includegraphics[width=0.5\textwidth]{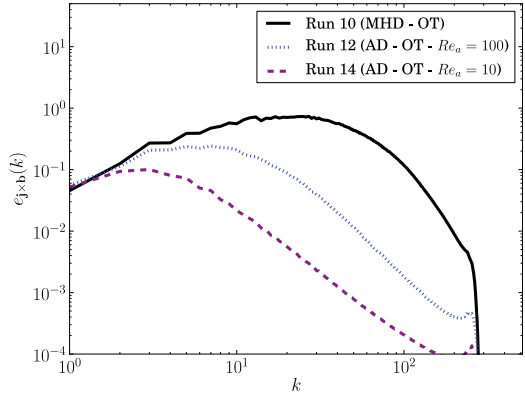}
  \caption{Power spectra of $\mathbf{j} \times \mathbf{b}$ for high
    resolution runs 10,12 and 14.  The field becomes force-free at
    small scales when the strength of the AD is increased.\label{fig:jxb-spectra}}
\end{figure}

The AD dissipation term, $Re_a^{-1} (\mathbf{j} \times \mathbf{b})^2$,
is proportional to the square of the Lorentz force.  It is hence
interesting to look at the influence of AD on the characteristics of
the Lorentz force, in comparison with the pure MHD case. Figure
\ref{fig:jxb-spectra} shows the spectrum of the Lorentz force for OT
runs 10, 12 and 14. The inclusion of AD has a significant effect on
the total power of the Lorentz force, reducing it importantly
especially in the dissipative range.  By contrast, as seen on figure
\ref{fig:spectra}, AD results in a deficit in the magnetic energy
spectrum (and hence the spectrum of the current vector) which is much
smaller in comparison to the deficit of the cross-product of these two
vectors (the Lorentz force), and is only present on intermediate
scales.

This is explained if AD has the effect of aligning the vectors
$\mathbf{j}$ and $\mathbf{b}$ \citep[as was also found in the
simulations of ][]{brandenburg-etal-1995}, with a stronger effect at
small scales.  To put it in an other way, AD leads the magnetic field
at small scales closer to a Lorentz force free configuration, where
the feedback of the magnetic field evolution on the velocity field
dynamics is weaker than for MHD. This was also found in the
simulations by \citet{brandenburg-zweibel-1995}.

Here we attempt to trace this tendency back to the evolution
equations.  We write $\mathbf{j}_{\rm \perp}$ the component of the
current vector $\mathbf{j}$: the double cross-product
$(\mathbf{j}\times\mathbf{b})\times\mathbf{b}$ can then be simply
written $b^2\mathbf{j}_{\rm \perp}$. This allows to write
\begin{equation}
  \label{eq:jperp}
  \left(\partial_t{\mathbf{j}}\right)_\mathrm{with AD}=\left(\partial_t{\mathbf{j}}\right)_\mathrm{without AD}+
  \nabla\times[\nabla\times(Re_a^{-1}b^2 \, \mathbf{j}_{\rm \perp})]
\end{equation}
which shows that in regions where $b^2$ is smooth enough, the effect
of AD is to diffuse out the component of the current perpendicular to
$\mathbf{b}$, and it does so faster at small scales like any
diffusion process. Hence AD brings the field closer to a force-free
state, more efficiently at small scales, except perhaps at the
smallest scales where $b^2$ varies and the behavior of equation
\eqref{eq:jperp} is less easy to predict.

\subsection{Spatial structure}

\begin{figure}
  \centering
  \includegraphics[width=0.5\textwidth]{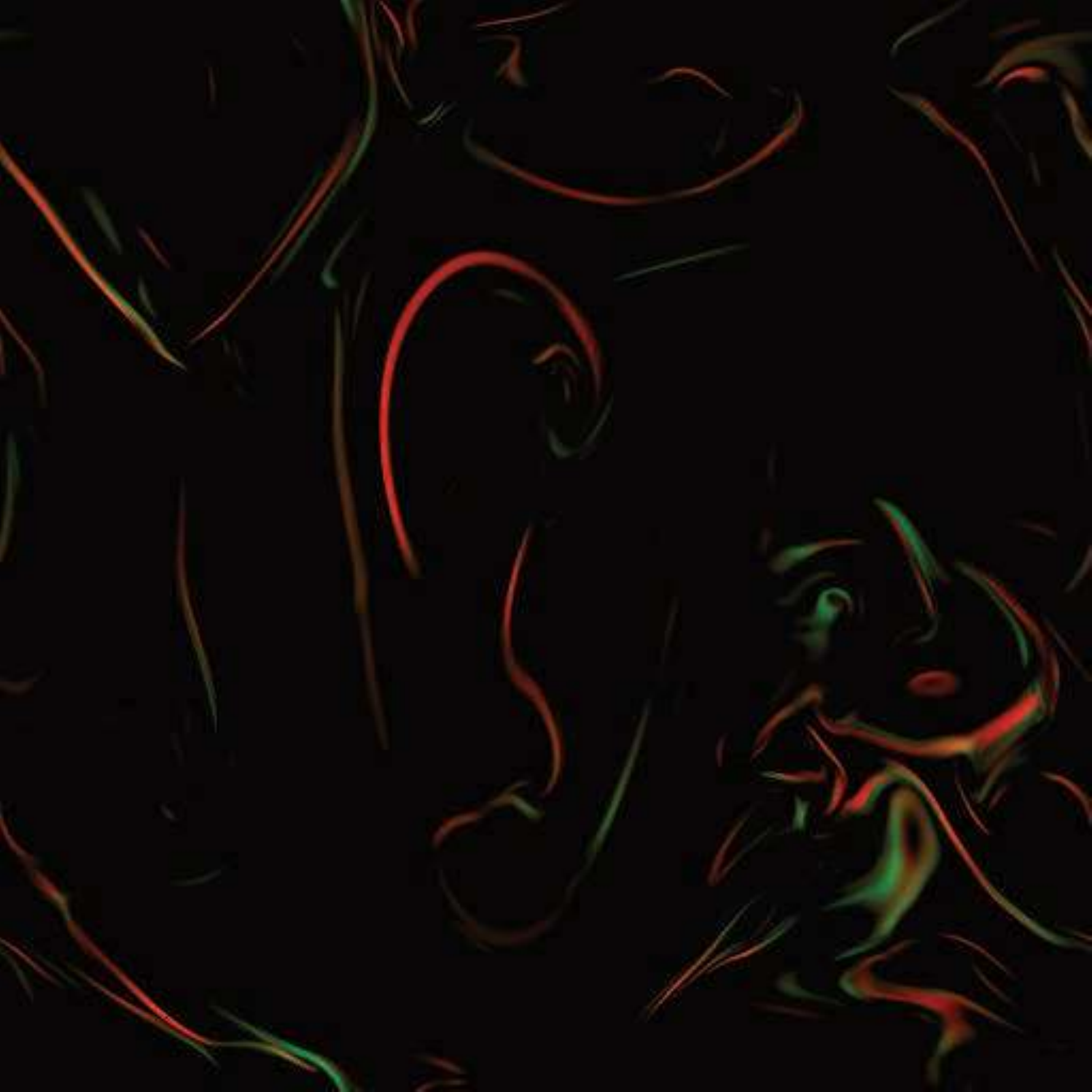}
  \caption{Color maps of a slice of the dissipation fields, run 10
    (MHD - OT). Red: Ohmic dissipation, green: viscous
    dissipation. All snapshots are taken at the peak of
    dissipation.\label{fig:rgb-mhd}}
\end{figure}

\begin{figure}
  \centering
  \includegraphics[width=0.5\textwidth]{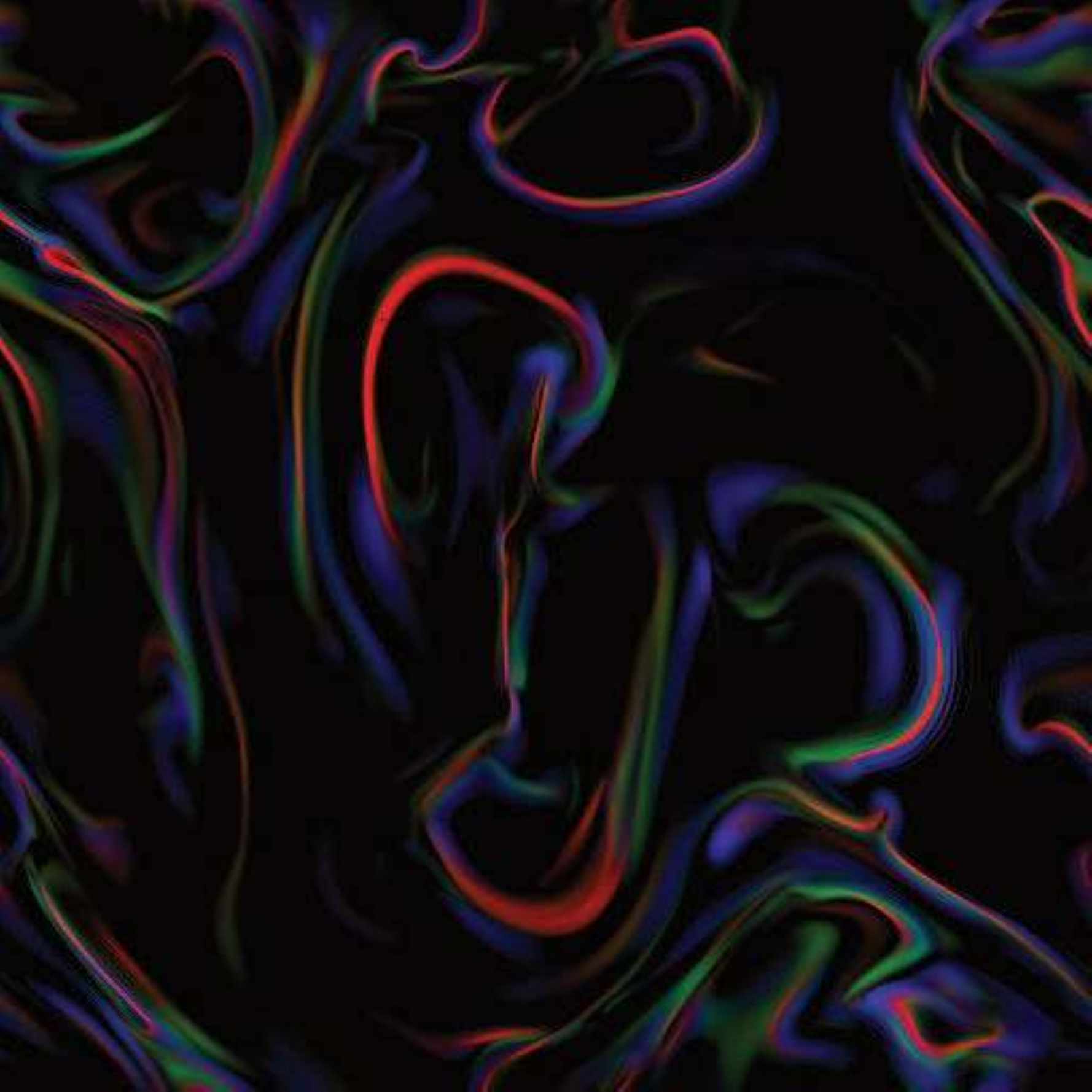}
  \caption{Same as Figure \ref{fig:rgb-mhd}, Run 12, (AD - OT, $Re_a =
    100$) with blue: ambipolar diffusion heating.\label{fig:rgb-ad-0.01}}
\end{figure}

\begin{figure}
  \centering
  \includegraphics[width=0.5\textwidth]{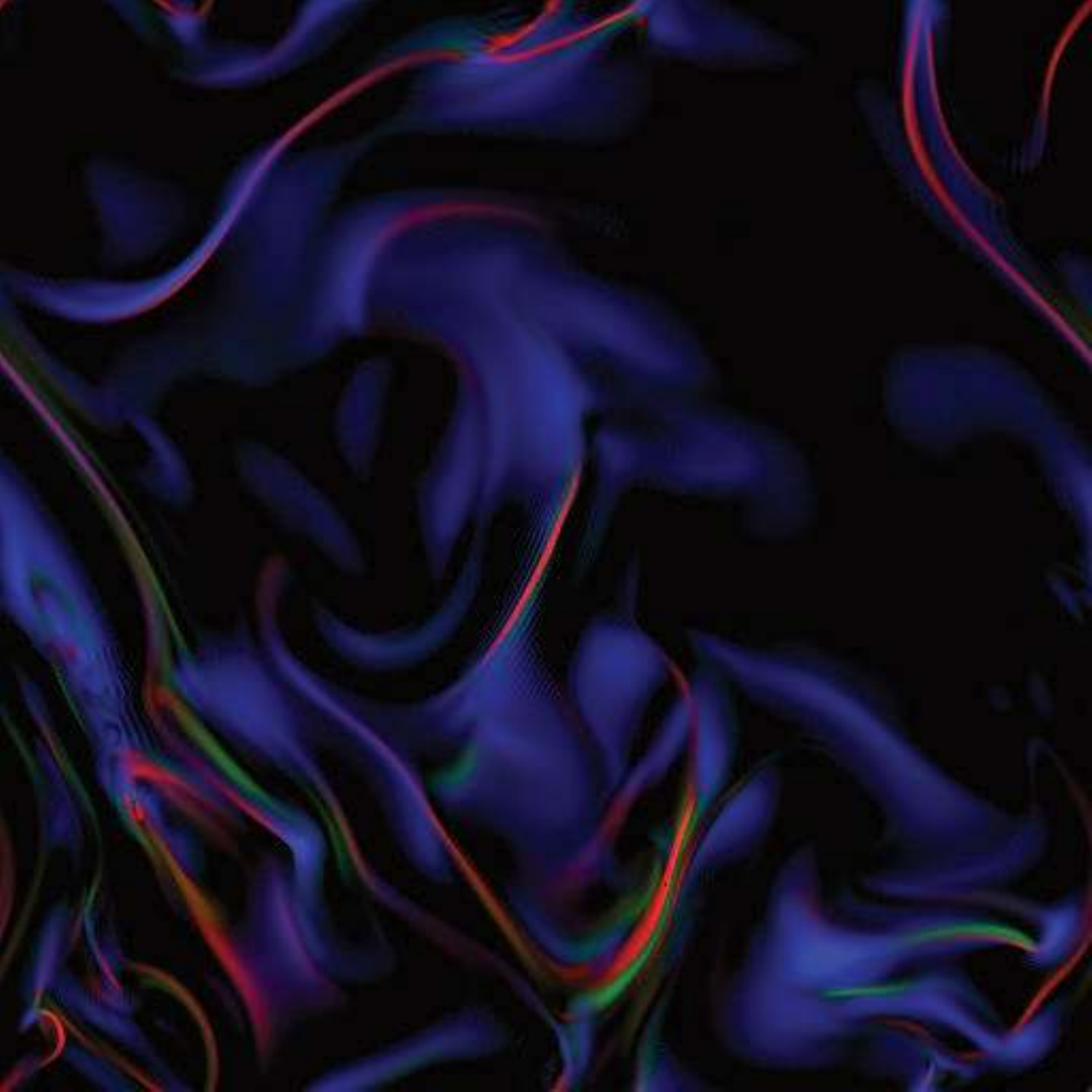}
  \caption{Same as Figure \ref{fig:rgb-mhd}, Run 14, (AD - OT, $Re_a =
    10$)  \label{fig:rgb-ad-0.1}}

\end{figure}

Next, we consider qualitatively the different contributions to the
bulk of the dissipation field.  For this purpose, each different
mechanism of dissipation (Ohmic, viscous and ambipolar diffusion) is
assigned to a color channel: Ohmic dissipation is assigned to the red
channel, viscous dissipation to the green channel and AD dissipation
to the blue channel. To emphasize the structures in the bulk of the
dissipation we first compute the total dissipation value
$\varepsilon_l$ below which 10\% of the heating occurs and the value
$\varepsilon_u$ below which 90\% of the dissipation occurs.  We
discard the pixels with total dissipation
$\varepsilon_t<\varepsilon_l$, we saturate the intensity of the pixels
with $\varepsilon_t>\varepsilon_u$ (while keeping their intrinsic
color) and we apply a logarithmic scaling for the intensity in between
these two thresholds.  The color of each channel is hence given by the
ratio of each type of dissipation to the total dissipation
\begin{equation*}
  \begin{split}
    \mathrm{Red} &= \frac{\varepsilon_o}{\varepsilon_t} ~ \mathrm{I}\\
    \mathrm{Green} &= \frac{\varepsilon_v}{\varepsilon_t} ~ \mathrm{I}\\
    \mathrm{Blue} &= \frac{\varepsilon_a}{\varepsilon_t} ~ \mathrm{I}
  \end{split}
\end{equation*}
with the intensity $\mathrm{I}$ given by
\begin{equation*}
  \mathrm{I} = \left\{
    \begin{array}{lll}
      0  & \mbox{if}  ~\varepsilon_t < \varepsilon_{l}\\
      \frac{\log(\varepsilon_t)-\log(\varepsilon_{l})}
      {\log(\varepsilon_{u})-\log(\varepsilon_{l})} &\mbox{if}  
      ~\varepsilon_{l} \leq
      \varepsilon_t \leq \varepsilon_{u}\\
      1 & \mbox{if } \varepsilon_{u} < ~\varepsilon_t 
    \end{array}
  \right.
\end{equation*}
The color maps of a slice through the dissipation fields are shown in
figures \ref{fig:rgb-mhd}-\ref{fig:rgb-ad-0.1} for all high-resolution
runs with the ABC initial conditions.

In the pure MHD case (Figure \ref{fig:rgb-mhd}), viscous and Ohmic
dissipation are in general concentrated on thin sheets: a slice by
slice inspection of the full cube reveals continuously evolving
filaments at the intersection between the sheets and the plane of the
slice. The length of the sheets is comparable to the integral length
scale \eqref{eq:ils} while their thickness is comparable to the
dissipation scale \eqref{eq:etad}.

The case of AD MHD (figures
\ref{fig:rgb-ad-0.01},\ref{fig:rgb-ad-0.1}) is similar in the sense
that viscous and Ohmic dissipation are again concentrated on thin
current sheets, although these sheets are fewer in number (more
connected), and the voids of low dissipation between them are smaller.
AD dissipation is significantly more diffuse than both Ohmic and
viscous dissipation, and is concentrated on much thicker
structures. The thickness of the AD dissipation structures seems to
coincide with the AD heating length $\ell_a^*$ measured on the power
spectra.  In some cases the structures of strong AD dissipation are
seen to surround the sheets of Ohmic or viscous dissipation: AD
sandwiches Ohmic dissipation, much like in the
\citet{brandenburg-zweibel-1994} picture.  This is probably also seen
on the left panel of Figure 7 of \citet{padoan-etal-2000} where the
structures of AD dissipation often go in pairs, except Ohmic
dissipation is absent from their simulations, so reconnection proceeds
only through numerical truncation errors.  Between figure
\ref{fig:rgb-mhd} and Figure \ref{fig:rgb-ad-0.01}, there is little
difference in the spatial structure of the dissipation field, with
similar size, shape and position for each sheets of dissipation.

The above color maps reveal only a small fraction of the pixels have
mixed colors (such as cyan, yellow or purple). This suggests that the
dissipative structures of different nature (Ohmic, viscous or AD) are
well separated.

\section{Intermittency and structures of high dissipation}

\subsection{Structure functions}

Figures \ref{fig:rgb-mhd}-\ref{fig:rgb-ad-0.1} suggest that the
dissipation field is not smoothly distributed in space, but is
characterized by a high degree of intermittency with regions of
extreme dissipation values alternating with relatively quiescent
regions.  The intermittent distribution of the energy dissipation rate
is expected to have an impact on the structure functions of the
velocity and the magnetic field. A longitudinal velocity structure
function of order $p$ is the $p$-th moment of the longitudinal
increment of the velocity field
\begin{equation*}
  S^u_p(r) = \langle (\delta u_\parallel(r))^p \rangle,\quad\delta 
  u_\parallel(r)=(\mathbf{u}(\mathbf{x}+\mathbf{r})-
  \mathbf{u}(\mathbf{x}))\cdot \hat{\mathbf{r}}
\end{equation*}
where $\hat{\mathbf{r}}$ is the unit vector in the direction of
$\mathbf{r}$. The definition for the structure function of the
magnetic field $S^b_p(r)$ is completely analogous. According to
standard \cite{kolmogorov-1941} phenomenology (hereafter K41), in the
inertial range the structure functions exhibit power--law scaling
\begin{equation*}
  S^u_p(r) \propto r^{\zeta^u_p},
  \qquad l_d \ll r \ll L
\end{equation*}
where the exponents $\zeta^u_p$ vary linearly with $p$, $\zeta^u_p =
p/3$. Intermittency considerations \citep{frisch-1995} lead to
deviations from the linear K41 prediction. A particularly successful
model of intermittency that was introduced for hydrodynamic turbulence
by \cite{she-leveque-1994} and generalized for MHD turbulence by
\cite{politano-pouquet-1995} predicts
\begin{equation*}
  \zeta^{GSL}_p(g,C) = \frac{p}{g} \left(1-\frac{2}{g}\right)+C\left(
    1-\left(1-\frac{2}{g C}\right)\right)^{\frac{p}{g}}
\end{equation*}
where $g$ is the inverse of the inertial range scaling exponent of the
velocity increment
\begin{equation*}
  \delta u_\parallel(r) \propto r^{1/g}
\end{equation*}
and $C$ is the co-dimension of the dissipative structures, $C=2$ for
filaments and $C=1$ for sheets in three space dimensions. K41
phenomenology predicts $g=3$, while Iroshnikov-Kraichnan (IK) MHD
phenomenology \citep{iroshnikov-1964,kraichnan-1965} predicts $g=4$.

\begin{figure}
  \centering
  \includegraphics[width=0.8\textwidth]{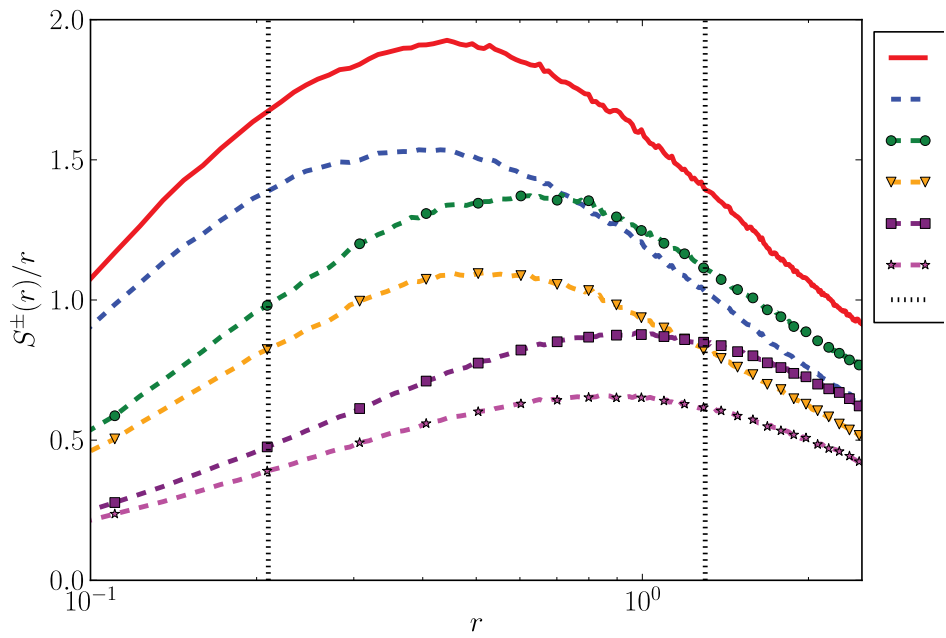}
  \caption{Compensated plot $S_z^{\pm}(r)/r$ for the OT runs 10,12 and
    14.\label{fig:sz-ot}}
\end{figure}

In order to estimate the structure function exponents from numerical
data, we used the extended self-similarity (ESS) method introduced by
\cite{benzi-et-al-1993}. The exponents were estimated by computing the
logarithmic slope
\begin{equation*}
  {\zeta^u_p} = \frac{d \log S_p^u(r)}{d \log S_*(r)}
\end{equation*}
where $S_*(r)$ is a structure function whose scaling behavior is known
from theory. In the case of MHD turbulence, $S_*(r)$ is given by the
two functions
\begin{equation*}
  S_z^{\pm}(r) = \langle \delta z_\parallel^{\mp} (\delta \mathbf{z}^{\pm})^2 \rangle
\end{equation*}
where $\delta \mathbf{z}^{\pm}$ is the increment of the Els\"asser
fields $\mathbf{z}^{\pm}=\mathbf{u} \pm \mathbf{b}$. Starting from the
MHD equations and assuming statistical homogeneity, isotropy and
stationarity, one can derive analytically
\begin{equation}
  \label{eq:fourthirds}
  S_z^\pm(r) = - \frac{4}{3} \langle \varepsilon^{\pm} \rangle r
  \qquad l_d \ll r \ll L
\end{equation}
where $\varepsilon^{\pm}$ are the dissipation rates of
$(\mathbf{z}^{\pm})^2$ \citep{politano-pouquet-1998-b}. The linear
scaling of $S_z^{\pm}(r)$ in the inertial range is confirmed
approximately by the flattening seen in Figure \ref{fig:sz-ot} which
shows compensated plots. Although the derivation of the law
\eqref{eq:fourthirds} is not proven in the case of AD MHD, the linear
scaling of $S_z^{\pm}(r)$ with $r$ is not further from linearity in
comparison to pure MHD.

\begin{figure}
  \centering
  \includegraphics[width=0.8\textwidth]{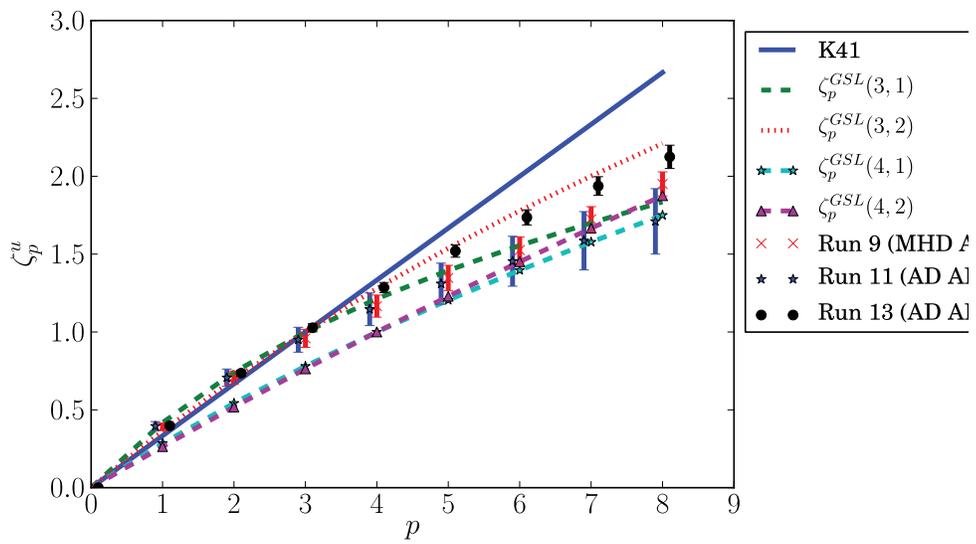}
  \caption{ESS velocity field structure function exponents for ABC
    runs 9-11-13.\label{fig:ess-vel-abc}} 
\end{figure}

\begin{figure}
  \centering
  \includegraphics[width=0.8\textwidth]{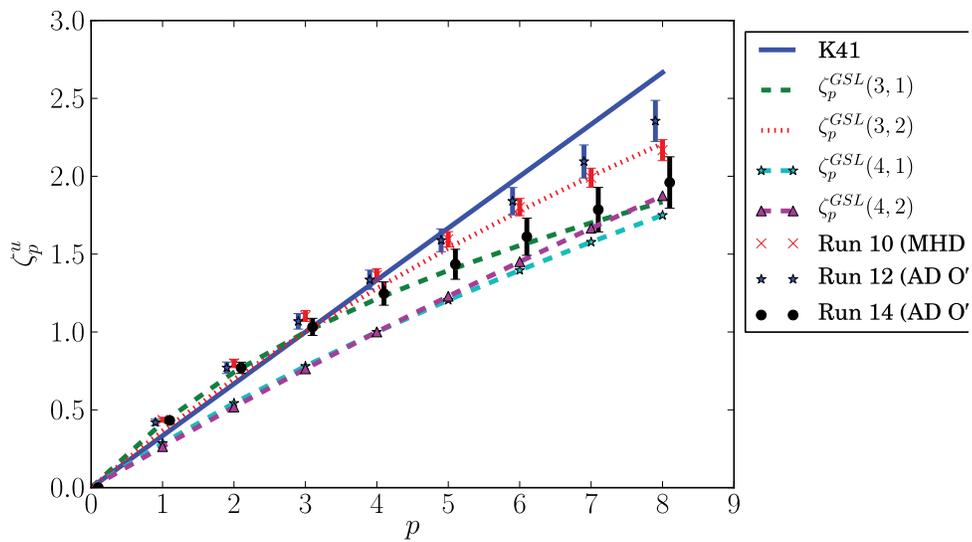}
  \caption{ESS velocity field structure function exponents for OT runs
    10-12-14.\label{fig:ess-vel}} 
\end{figure}

\begin{figure}
  \centering
  \includegraphics[width=0.8\textwidth]{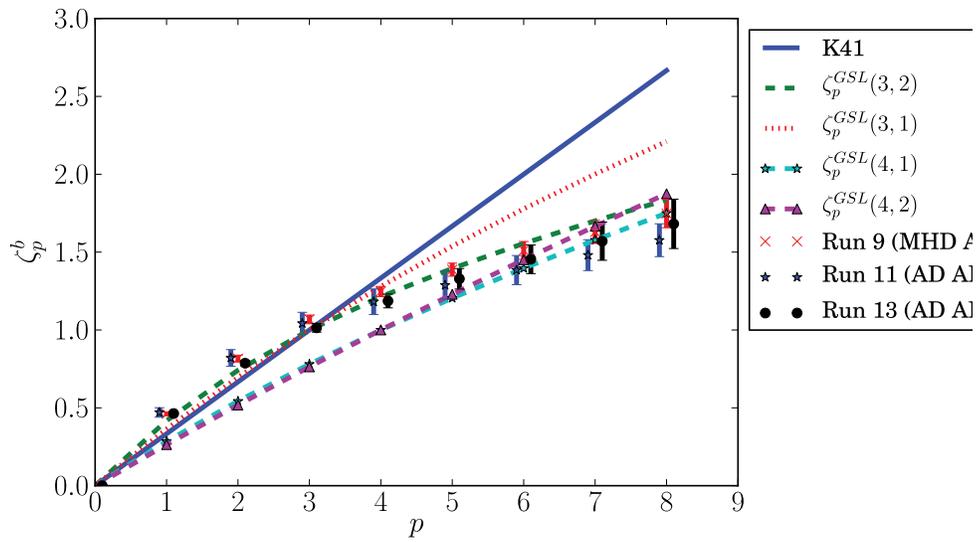}
  \caption{ESS magnetic field structure functions for ABC runs
    9-11-13.  \label{fig:ess-mag-abc}}
\end{figure}

\begin{figure}
  \centering
  \includegraphics[width=0.8\textwidth]{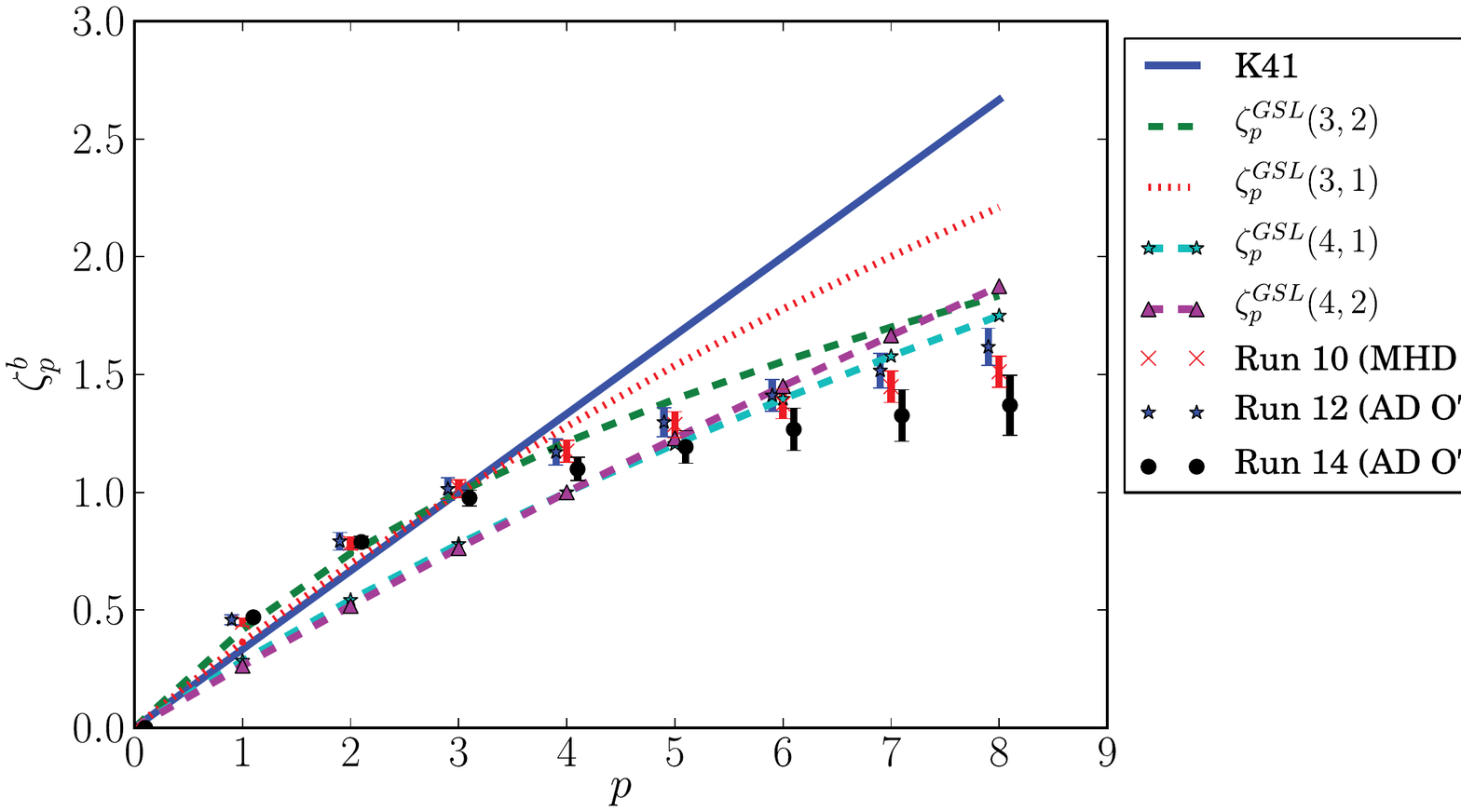}
  \caption{ESS magnetic field structure functions exponents for OT
    runs 10-12-14.  \label{fig:ess-mag}}
\end{figure}

The velocity and magnetic field structure function exponents up to
order $8$, calculated for high resolution runs 9-14 using extended
self-similarity are shown in figures
\ref{fig:ess-vel-abc}-\ref{fig:ess-mag}. We see a departure from the
linear prediction of K41, a sign of intermittency, especially in the
case of the magnetic field. The structure function exponents follow
closely, but not exactly, the predictions of the generalized She \&
Lev\^eque model with a scaling parameter $g=3$ and a co-dimension for
the structures of high dissipation around $C=1$ (with exceptions at
$C=2$ and some below $C=1$). This fact suggests that in the inertial
range $\delta u(r)$ and $\delta b(r)$ are proportional to $r^{1/3}$,
which is the prediction of K41 phenomenology, rather than $r^{1/4}$ as
predicted by IK phenomenology. The value $C=1$ of the co-dimension
suggests that the structures of high dissipation in the inertial range
are sheet-like, in accordance with figures
\ref{fig:rgb-mhd}-\ref{fig:rgb-ad-0.1} and the analysis of section
\ref{sect:extraction}. The velocity exponents for runs 9, 10 and 12
hint towards $C=2$ (filaments) and the magnetic exponents for the OT
runs exhibit a greater degree of intermittency. This indicates that
even though the exponents appear to follow generalized She \&
Lev\^eque models, the model's phenomenology probably does not subtend
the dissipation in our simulations.

The effect of AD on intermittency is not easy to discern from these
scaling exponents. In the case of the ABC initial condition, the
deviation from the K41 values is larger for both the velocity and the
magnetic field. The situation is reversed in the case of the OT
initial condition, where the pure MHD fields appear to be more
intermittent.

\subsection{Extraction of structures \label{sect:extraction}}

Following \cite{uritsky-et-al} (hereafter UR10), we implemented an
algorithm for the extraction of structures of high dissipation. In the
problem considered, there are three types of local dissipation rates:
the viscous dissipation rate,
\begin{equation}
  \label{eq:viscdiss}
  \varepsilon_v = \frac{Re^{-1}}{2} \sum_{i,j=1}^{3}\left(\partial_i u_j + 
    \partial_j u_i \right)^2
\end{equation}
the Ohmic dissipation rate
\begin{equation}
  \label{eq:ohmdiss}
  \varepsilon_o = Re_m^{-1}~\mathbf{j}^2 = Re_m^{-1} \left( \nabla \times \mathbf{b} \right)^2
\end{equation}
and the AD dissipation rate
\begin{equation}
  \label{eq:addiss}
  \varepsilon_a = Re_a^{-1}\left(\mathbf{j}\times\mathbf{b}\right)^2 \mbox{.}
\end{equation}
A structure of high dissipation is defined as a connected set of
points $\mathbf{x}$ where
\begin{equation}
  \label{eq:extr}
  \varepsilon(\mathbf{x}) > \langle \varepsilon \rangle + j \sigma_\varepsilon\mbox{, $j=$1,2 or 3.}
\end{equation}
In the above relation, $\varepsilon$ can be any of the three
dissipation rates or the total dissipation rate $\varepsilon_t =
\varepsilon_v + \varepsilon_o + \varepsilon_a$ (or $\varepsilon_t =
\varepsilon_v + \varepsilon_o$ in the case AD is absent). $\langle
\varepsilon \rangle$ is the spatial mean value of the dissipation rate
and $\sigma_\varepsilon$ its standard deviation. For example, the
three thresholds we use on total dissipation for run 12 are displayed
on figures \ref{fig:pdf} and \ref{fig:cdf} over the PDF and the CDF of
the total dissipation.

\begin{figure}
  \centering
  \includegraphics[width=0.5\textwidth]{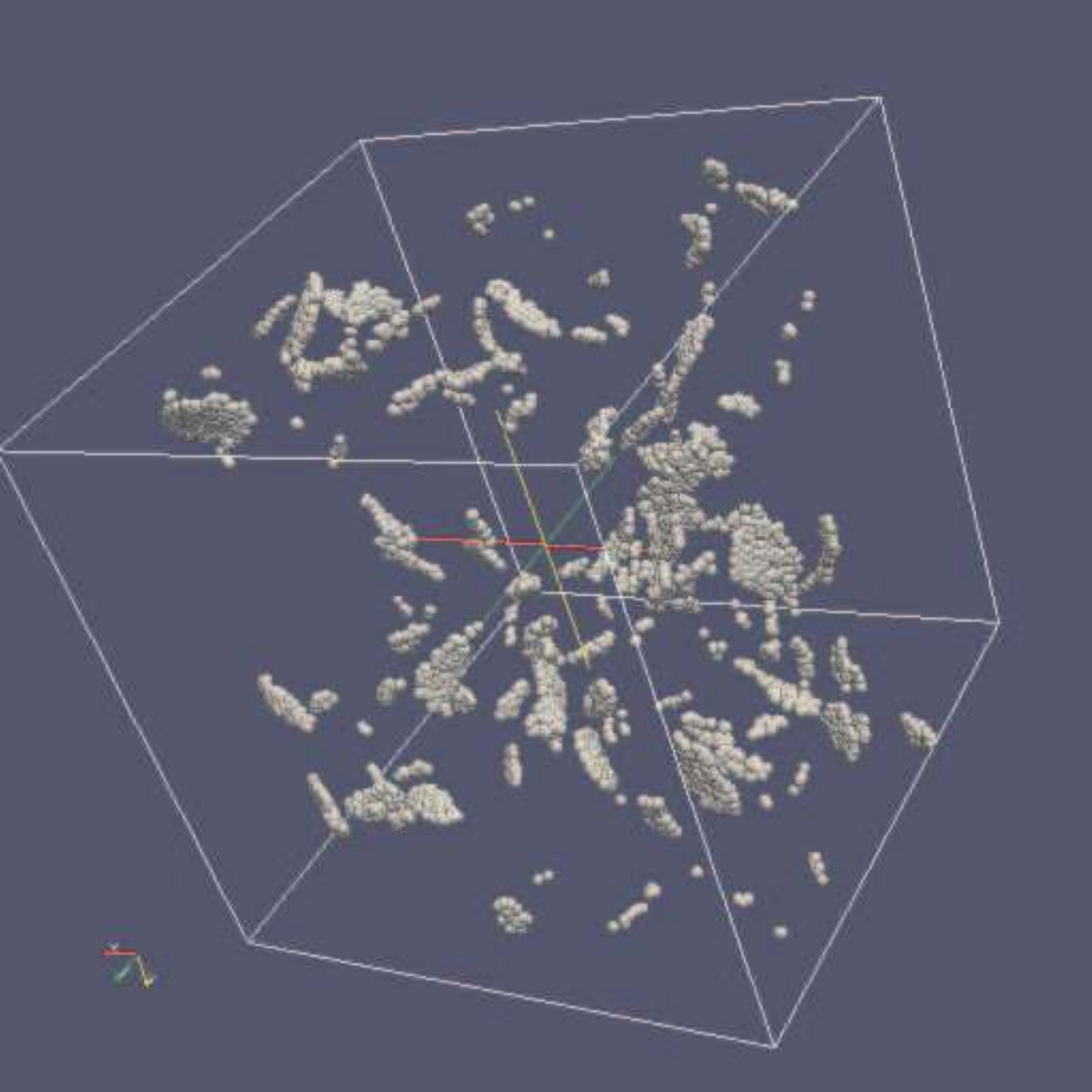}
  \caption{Inertial range structures extracted from the dataset
    corresponding to the peak of dissipation of Run 12 (AD - OT).
    They are defined as connected sets of points having values of the
    total dissipation three standard deviations above the mean
    value. Each little sphere has a 2-pixels diameter, ie: about the
    size of the viscous (or equivalently Ohmic) dissipation
    length.\label{fig:struct-inertial}} 
\end{figure}

\begin{figure}
  \centering
  \includegraphics[width=0.5\textwidth]{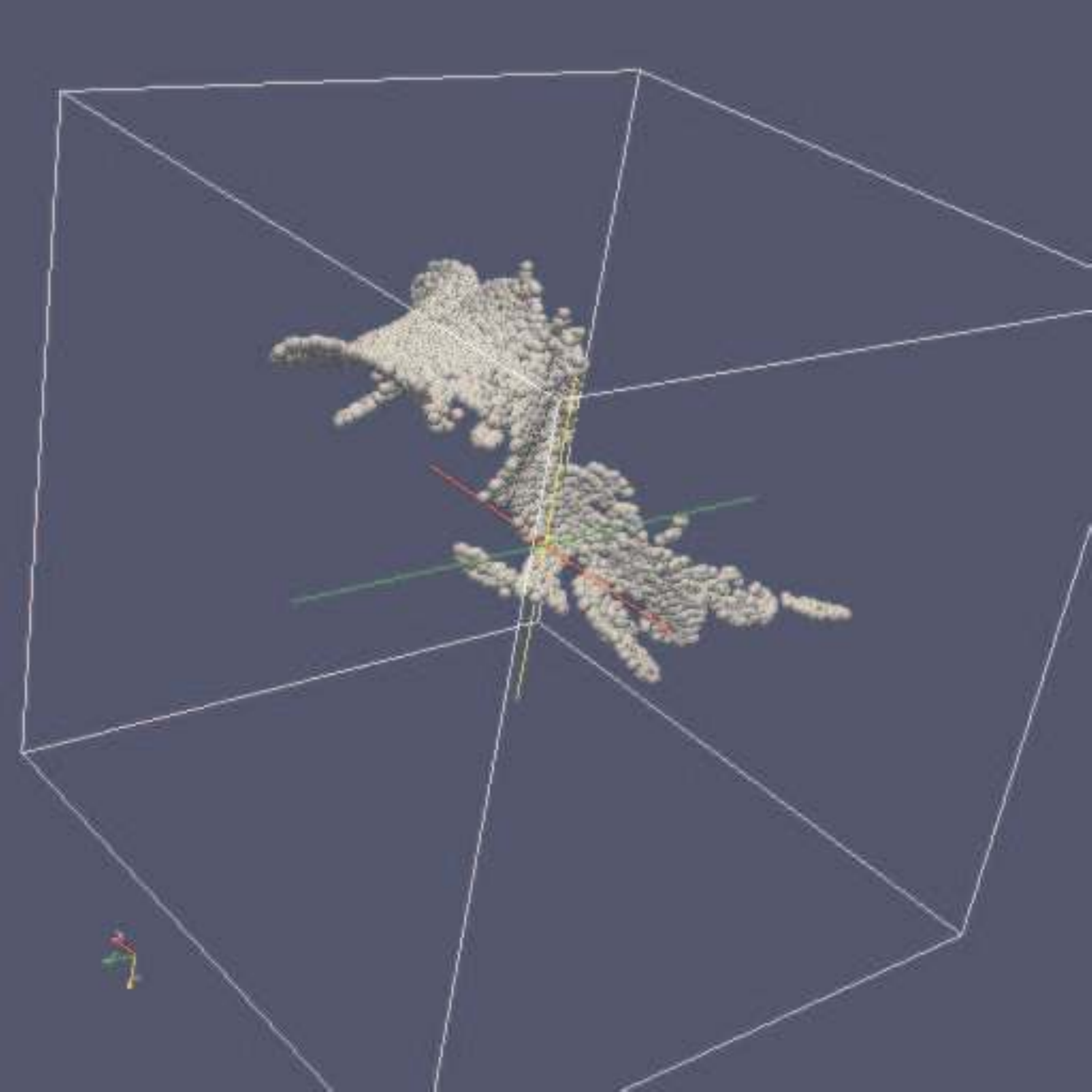}
  \caption{One of the largest structures extracted from run
    12\label{fig:struct}} 
\end{figure}

The algorithm is capable of isolating the structures of high
dissipation so that a statistical analysis of their geometric and
dynamical characteristics can be performed. The extracted structures
are generally sheet-like, as can be seen in figure
\ref{fig:struct-inertial}, where all structures extracted from run 12
(AD - OT) whose characteristic linear sizes $L_i$ (see below) lie in
the inertial range are shown. An example of a more complex structure
can be seen in Figure \ref{fig:struct}. This structure is one of the
largest extracted from this dataset.  It is sheet-like, with its
length larger than the integral length scale \eqref{eq:ils} and its
thickness is comparable to the dissipation scale \eqref{eq:etad}. It
is overall characterized by a high degree of geometrical complexity.

After the extraction of the structures, the following quantities were
computed for statistical analysis
\begin{subequations}
  \label{eq:quants}
  \begin{align}
    L_i &= \delta \max_{m,l \in \Lambda_i} \lvert \mathbf{r}_m -
    \mathbf{r}_l \rvert\\
    L_{\{x,y,z\}i} &= \delta \max_{m,l \in \Lambda_i} \lvert
    \{x,y,z\}_m -
    \{x,y,z\}_l \rvert\\
    R_i &= \sqrt{L_{xi}^2 + L_{yi}^2 + L_{zi}^2}\\
    V_i&=\delta^3 \sum_{m\in\Lambda_i}1\\
    A_i&=\delta^2 \sum_{\substack{m \in \Lambda_i \\ M(m) \not\subset \Lambda_i}} 1\\
    P_i &= \delta^3 \sum_{m \in \Lambda_i} \varepsilon(\mathbf{r}_m)
  \end{align}
\end{subequations}
In the above definitions, $\Lambda_i$ is the i-th structure, $m$ its
$m$-th point, $M(k)$ the set of 26 neighbors of the point $m$ and
$\delta=2\pi/N$ is the grid spacing. $L_i$ is the characteristic
linear scale of the structure, $L_{\{x,y,z\}i}$ is the linear scale of
its projection on the three axes, $R_i$ is the characteristic linear
scale of the smallest volume embedding the whole structure, $V_i$ is
its volume, $A_i$ its surface and $P_i$ the volume-integrated
dissipation rate.  In the next section, all statistical quantities are
computed on the sample of different extracted structures. Note that
the definition of $R_i$ makes it dependent of the orientation of the
structure (its value can change by tilting slightly the axis of the
domain).

In order to validate our extraction procedure, we implemented it in
two different ways. First, we implemented the same algorithm as UR10
(ie: recursive, using breadth-first search as explained in
UR10). Second, we implemented a non-recursive algorithm: we parse the
whole cube to find a pixel above threshold which is not yet included
into a structure ; we tag it and we scan the whole cube several times
to tag the neighboring pixels of this growing seed which happen to be
above threshold, until we find no new pixel to attach to this
structure ; finally we reiterate to find another pixel not yet
included in a structure until we fail to find any new pixel above
threshold. The second algorithm is much more CPU time consuming, but
keeps the memory usage constant and is easier to implement.  We
checked that both algorithms identify strictly the same structures for
the two implementations in the low resolution cases.

\begin{figure}
  \begin{center}
    \centering
    \includegraphics[width=0.5\textwidth]{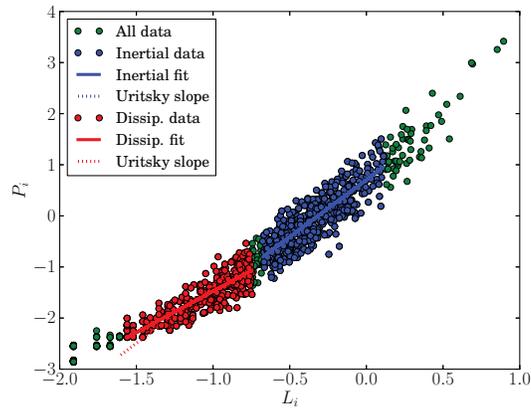}
    \caption{Scaling relations $P_i \propto L_i^{D_P}$ from Run 12 (AD
      - OT), at the peak of dissipation, with a threshold of two
      standard deviations above mean value. 
      The dotted line shows the effect of adopting the slope found by
      \citet{uritsky-et-al} instead of our own slope.\label{fig:scaling-L}}
  \end{center}
\end{figure}

Like in UR10, all the above quantities are found to exhibit power-law
scaling with respect to structure linear size $L_i$, with different
scaling behavior in the inertial and dissipative ranges. The quantity
$X_i$, which could be any of $L_i,R_i,V_i,A_i,P_i$ scales as
\begin{equation*}
  X_i \propto L_i^{D_X}
\end{equation*}
with different scaling exponents $D_X$ in the inertial and dissipative
ranges, while the pdf of $X_i$ scales as
\begin{equation*}
  \mathcal{P}(X_i) \propto X_i^{-\tau_X}
\end{equation*}
with different scaling exponents $\tau_X$ in the inertial and
dissipative ranges. As an example, the scaling relations $P_i \propto
L_i^{D_P}$ and $\mathcal{P}(P_i) \propto P_i^{-\tau_P}$ are shown in
figures \ref{fig:scaling-L} for the structures extracted from Run 12
(AD - OT), at the peak of dissipation, with a threshold of two
standard deviations above mean value. The limits of the inertial and
dissipation ranges are also shown (as used by UR10, see section
\ref{ranges}).  The upper limit of the dissipation range is just below
the lower limit of the inertial range, while the lower limit of the
dissipation range is $\sim 2$ times the numerical resolution.

\subsection{Comparison with UR10}

In this section we compare the results of the statistical analysis of
structures of high dissipation with those of UR10. These authors
consider pure MHD and study the structures of high Ohmic dissipation
or high enstrophy
\begin{equation*}
  \varepsilon_\omega = Re^{-1} \boldsymbol{\omega}^2,\qquad 
  \boldsymbol{\omega}=\nabla \times \mathbf{u}
\end{equation*}

\begin{figure}
  \centering
  \includegraphics[width=0.8\textwidth]{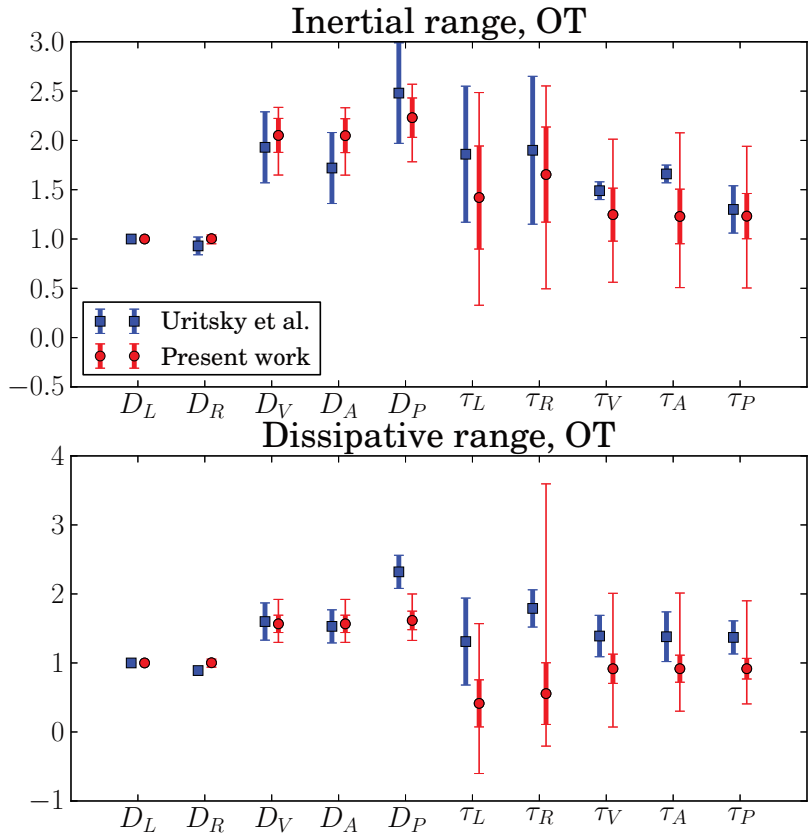}
  \caption{Scaling exponents for structures extracted based on the
    Ohmic dissipation (red circles), comparison with results of UR10
    (blue squares). Upper panel: Inertial range exponents for our Run
    10 (MHD-OT) compared with the corresponding Run III of UR10. Lower
    panel: Dissipative range exponents for the same runs. Thick error
    bars are three-sigma error brackets. Thin red error bars estimate
    the systematics related to the choice of the boundaries of the
    inertial and dissipative ranges as well as the bin size for the
    definition of the pdfs: see text for details. \label{fig:exps-uritsky}}
\end{figure}

The results for the scaling exponents are shown in figure
\ref{fig:exps-uritsky}, for the case of the OT initial condition. The
case of the ABC initial condition (not shown) is similar. The
exponents are calculated from run 10 which in terms of initial
condition and Reynolds number is similar with run III of UR10. As in
the present paper, the snapshot analyzed is on the peak of
dissipation. The structures of high dissipation are defined as
connected sets of points having values of the Ohmic dissipation two
standard deviations above the mean value. We keep the same definitions
as UR10 for the inertial and dissipative ranges (cf. section
\ref{ranges}).

Figure \ref{fig:exps-uritsky} shows that the agreement between our
results and those of UR10 is not completely satisfactory. Although
most three-sigma error bars are compatible and the errors are on the
same order, there remains systematic differences, especially for our
pdf exponents which appear to yield shallower pdfs than UR10. The
least square method used by UR10 to estimate the slope of power-law
pdfs is known to introduce some bias and the maximum-likelihood
estimate method (MLE) should be used instead
\citep[see][]{clauset-etal-2009}. We computed the exponents with the
MLE, and found them to be very close to our least-square values.  We
turned to explore the effects of some systematics due to the
uncertainty on the boundaries of the inertial and dissipative range
and the size of the bins used to produce the pdfs.  We varied randomly
these parameters with factors in an octave centered on their initial
chosen values. The excursion of the resulting three-sigma error bars
over a thousand of such realizations are plotted as thin red error
bars on Figure \ref{fig:exps-uritsky}. A slight displacement of the
inertial or dissipative range boundaries incorporates (or leaves out)
new data near the edge of the fitting intervals, where their leverage
on the fitted slope is quite important. The resulting figure shows
that such systematics can account for nearly all discrepancies with
respect to UR10, except for the correlation between the total
dissipation of structures and their linear size. We illustrate the
corresponding discrepancy on Figure \ref{fig:scaling-L} where the
dotted red line shows the slope followed by UR10 data: this line
clearly falls below our data at small scales.  This suggests that our
smallest structure have a higher value of dissipation.  This may
perhaps be traced back to the slightly more refined dealiasing rule
which we use. The picture is the same for the ABC runs, except for the
error bar 0.02 on $D_A$ in Run I of UR10 (see their Table II) which is
probably a typo as the value they quote does not correspond to the
scatter displayed in their Figure 3.

To validate further our results, we compared the computed scaling
exponents for the dataset corresponding to the temporal peak of total
dissipation of the pure MHD run with ABC or OT initial condition (Run
9 and 10) with those computed from a snapshot taken one macroscopic
eddy turnover time later, in the decay period of the turbulence. In
agreement with the results of UR10, we find no statistical difference
between the two snapshots (one-sigma error bars are compatible).
Similarly, we compared the exponent values computed based on
$\mathbf{j}^2$ with those computed based on $\boldsymbol{\omega}^2$
for the dataset corresponding to the peak of total dissipation of the
same runs (Runs 9 and 10). Again the exponents were seen to be
compatible within one-sigma error bars, as in UR10.

 
\subsection{Structures based on total dissipation}

In this section we focus on the statistical analysis of structures
extracted based on the total dissipation
$\varepsilon_t=\varepsilon_o+\varepsilon_v+\varepsilon_a$ for both
pure MHD and AD MHD.  The analysis based on the total dissipation is
more relevant to the heating of the ISM because all three different
types of dissipation can be important heating agents. AD has an
additional specificity because the ion-neutral drift increases the
effective temperature of the chemical reactions, but we do not
consider this yet in the present work. In the following, we will note
$D_X$ for the linear size exponents and $\tau_X$ for the probability
exponents of a characteristic $X$.  All the structures discussed in
this section are defined as connected sets of points having values of
the total dissipation two standard deviations above the mean value.

\begin{figure}
  \begin{center}
    \centering
    \includegraphics[width=0.5\textwidth]{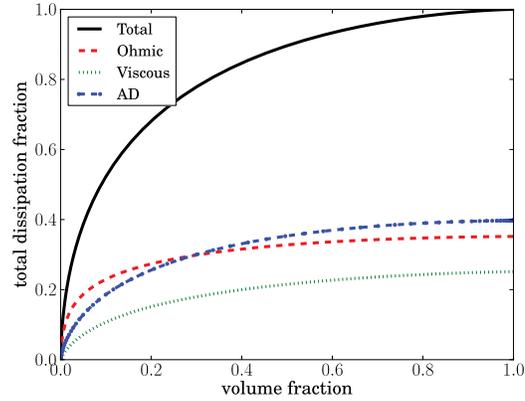}
    \caption{We look at the subset of pixels above a given threshold
      of total dissipation in run 12 (AD-OT, $Re_a=100$) at the
      dissipation peak.  For each value of the threshold, we plot the
      fraction of the total energy dissipation on this subset versus
      the volume of this subset (black curve). We also give the
      fraction of the total dissipation on this subset for each nature
      of dissipation (red: Ohmic, green: viscous, blue:
      AD).\label{fig:diss-vol}}
  \end{center}
\end{figure}

In Table \ref{tab:structs} we present the results of the structure
extraction algorithm. The relative amount of dissipation contained by
all the detected structures does not depend much on the Reynolds
number, but the volume filling factor of the structures decreases as
the Reynolds number increases. The presence of AD results in fewer
detected structures than without AD (except for run 13). However, they
fill roughly the same volume fraction, thus AD structures tend to be
larger.  This difference between the number of different structures in
pure MHD and AD increases with the Reynolds number.  Figure
\ref{fig:diss-vol} gives a more detailed view of the fraction of total
dissipation contained in structures with a value above a given
threshold as a function of the volume fraction occupied by these
structures. The curve rises steeply near the origin, so that 30
percent of the dissipation in contained in less than 3 percent of the
total volume. The steepness at the origin is seen to be mainly due to
the Ohmic heating: this is in line with the original picture of
\citet{brandenburg-zweibel-1994} where AD forms sharp features in
which Ohmic dissipation is favored.

\begin{table}
  \begin{center}
      \begin{tabular}{|c || c | c | c | c | c | c |}
        \hline
        \# & N & $Re_\lambda$ & $Re_a$ & \# of structures & \% of volume & \% of total dissipation \\ 
        \hline
        1 & 128 & 189 & -  & 87 & 3.90 & 23.85  \\
        \hline
        2 & 128 & 210 & - & 54 & 3.81 & 27.29  \\
        \hline
        3 & 128 & 217 & 100 & 72 & 3.78 & 23.65  \\
        \hline
        4 & 128 & 228 & 100 & 34 & 3.90 & 26.91  \\
        \hline
        5 & 256 & 353 & - & 215 & 3.26 & 27.32  \\
        \hline
        6 & 256 & 377 & - & 233 & 3.26 & 26.80  \\
        \hline
        7 & 256 & 412 & 100 & 144 & 2.86 & 26.40  \\
        \hline
        8 & 256 & 444 & 100 & 153 & 2.98 & 26.53  \\
        \hline
        9 & 512 & 591 & - & 790 & 2.70 & 31.30  \\
        \hline
        10 & 512 & 604 & - & 1166 & 2.77 & 30.33  \\
        \hline
        11 & 512 & 756 & 100 & 375 & 2.17 & 29.21  \\
        \hline
        12 & 512 & 750 & 100 & 418 & 2.57 & 29.48  \\
        \hline
        13 & 512 & 640 & 10 & 1167 & 2.94 & 28.01  \\
        \hline
        14 & 512 & 927 & 10 & 378 & 1.75 & 22.00  \\
        \hline
      \end{tabular}
    \caption{Results of the structure extraction algorithm for all
      runs and structures defined as connected sets of points having a
      value of the total dissipation two standard deviations above
      mean value.\label{tab:structs}}
  \end{center}
\end{table}

\begin{figure}
  \centering
  \includegraphics[width=0.8\textwidth]{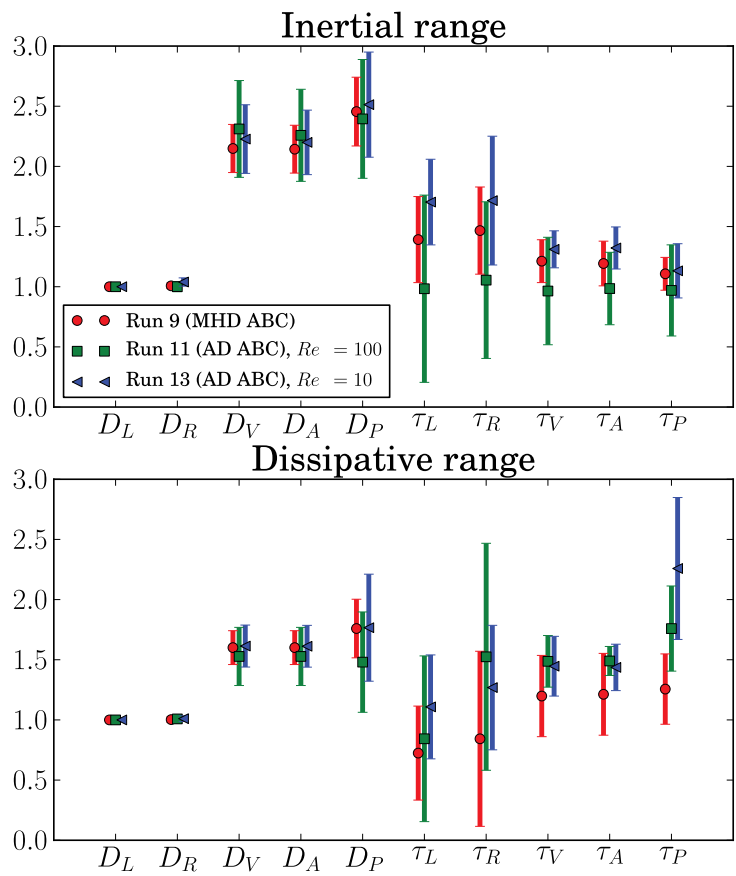}
  \caption{Comparison of scaling exponents with three-sigma error bars
    between pure MHD (red circles) and AD MHD (green squares
    $Re_a=100$ and blue triangles $Re_a=10$) - ABC Runs 9,11,13\label{fig:tot-abc}}
\end{figure}
    
\begin{figure}
  \centering
  \includegraphics[width=0.8\textwidth]{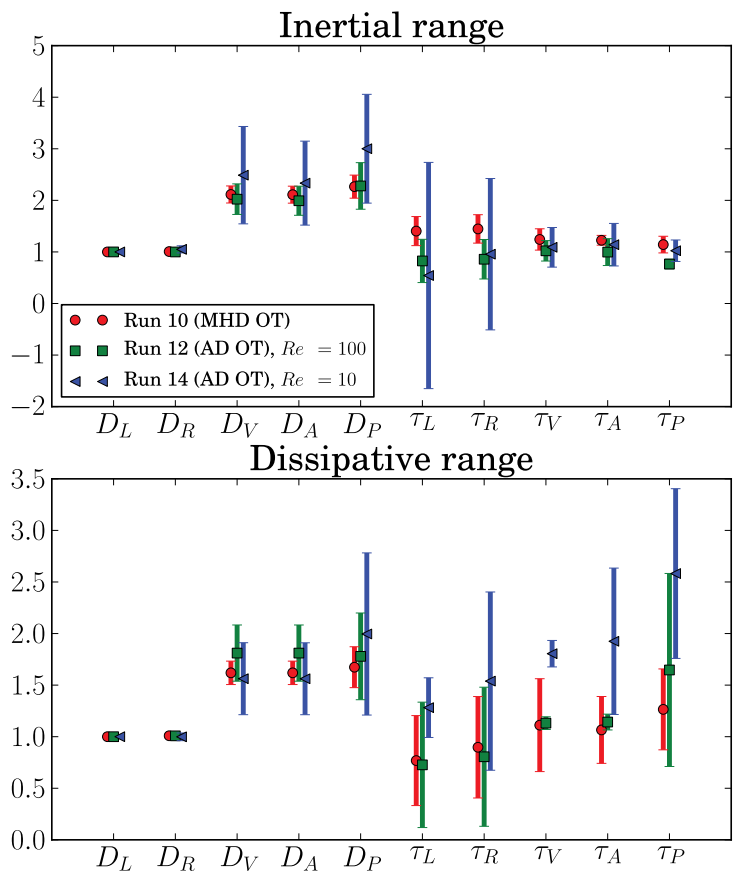}
  \caption{Same as Figure \ref{fig:tot-abc} but for the OT Runs
    10,12,14.\label{fig:tot-ot}}
\end{figure}
    

Figures \ref{fig:tot-abc} and \ref{fig:tot-ot} summarize all results
on the exponents for the total dissipation two standard deviations
above the mean. Although most three-sigma error bars are compatible,
in particular for the $D_X$ exponents which are unchanged with AD,
some systematic differences exist for the pdfs exponents $\tau_X$. The
pdfs exponents are in general steeper in the dissipation range: AD
seems to favor more fragmented structures in the dissipative
range. This confirms the tendency for more intermittency with AD that
was suggested by the structure functions analysis.  However, we see no
clear cut tendency in the inertial range. If we discard the strong AD
results ($Re_a=10$), the inertial range shows a behavior opposite
from the dissipative range (shallower pdfs slope, ie: larger
structures are favored when AD is present). However if we now look
only at the MHD runs and the $Re_a=10$ runs, the inertial range sees
no change in the $\tau_X$ exponents for the OT runs, but bigger
exponents for the ABC runs, in agreement with the dissipative range
and contrary to the $Re_a=100$ runs...

This complicated picture might perhaps not be genuine, as the huge
systematics experienced for the comparison with UR10 show. However,
the dependence of the intermittency statistics on the initial
conditions (seen both in the structure pdfs slopes and in the
structure functions exponents) points at their difference in magnetic
helicity content. In the OT initial condition, the initial value of
magnetic helicity is almost zero, and the equation of magnetic
helicity evolution is unchanged by the inclusion of the AD term. As
far as the effects of viscosity and resistivity are neglected, both
the pure MHD and AD MHD solution evolve under the same constraint of
zero magnetic helicity. A zero value for the magnetic helicity is an
important constraint because it implies statistical reflection
invariance, a property which the ABC runs will not share. In the ABC
initial condition, the constraint of very low cross-helicity is broken
by AD, which provides a source term in the cross-helicity equation.


\begin{figure}
  \centering
  \includegraphics[width=0.5\textwidth]{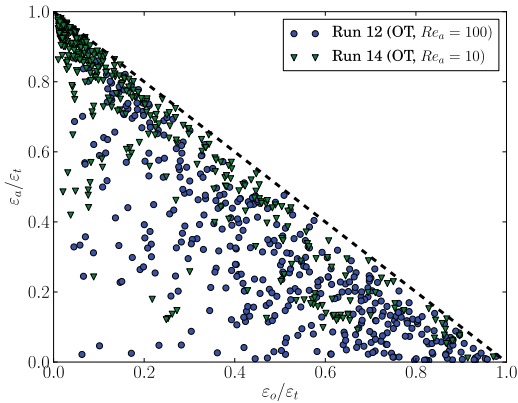}
  \caption{Scatter plots of the ratio
    $\varepsilon_{o}/\varepsilon_{t}$ versus the ratio
    $\varepsilon_{a}/\varepsilon_{t}$ for AD-OT runs 12 and 14.\label{fig:scatter-ot}}
\end{figure}
    

The above analysis does not give any information on the relative
amount of Ohmic, viscous and AD dissipation contained within each
structure. To answer this question, we show in figure
\ref{fig:scatter-ot} scatter plots of the ratio of total Ohmic
dissipation to total dissipation $\varepsilon_{o}/\varepsilon_{t}$
versus the ratio of total AD dissipation to total dissipation
$\varepsilon_{a}/\varepsilon_{t}$ in the two OT runs 12 and 14. In all
AD cases there is a tendency for the structures to cluster close to
the line of zero viscous dissipation (dashed line in figure
\ref{fig:scatter-ot}). This shows that within the structures of high
total dissipation, viscous dissipation is relatively less
important. The relative values of Ohmic and AD dissipation span
however the whole spectrum, in contrast to the impression given by our
RGB slices (see figures \ref{fig:rgb-mhd} to \ref{fig:rgb-ad-0.1}):
this could be a genuine difference between the extreme dissipation
events and the bulk of the dissipation shown on the RGB figures, or
the extraction algorithm of connected structures could merge nearby
sheets with different dissipation natures. We checked on a pixel by
pixel scatter plot similar to Figure \ref{fig:scatter-ot} that it is
indeed a genuine difference. For the strongest AD runs, though, the
intense dissipation structures tend to be predominantly due to AD
heating (top left corner in Figure \ref{fig:scatter-ot}).

\section{Concluding remarks}

We performed a 3D numerical study of the structures of high
dissipation in MHD turbulence, with the inclusion of ambipolar
diffusion. At the Reynolds numbers studied, the total dissipation due
to viscosity, resistivity and AD are of comparable magnitude.

Kinetic and magnetic energy spectra show that ambipolar diffusion
enhances the turbulent energy to small scales at the expense of
intermediate scales. This agrees with the idea of
\citet{brandenburg-zweibel-1994} that AD can sharpen the magnetic
gradients, but the effect is not strong enough to increase the total
Ohmic dissipation rate in our simulations. Previous authors
\citet{lietalI} and \citet{oishi-maclow-2006} have examined the case
of driven two-fluids compressible turbulence with a mean magnetic
field and they find different results: \citet{oishi-maclow-2006} find
no effect on the slope of the spectrum (for $Re_a=2.5$ and $Re_a=5$)
while \citet{lietalI} find that AD {\it steepens} it, although they
don't display the results for their runs with $Re_a=12$ and
$Re_a=120$, the only ones which have $\ell_a$ in the computed range of
scales as in our study. It should be noted that both these papers
neglect Ohmic diffusion, and rely on truncation errors only to
reconnect the field: the present work is the first to account for AD
in the presence of controlled Ohmic and viscous dissipation.  This is
important because the dissipation physics can be quite different from
the numerical dissipation as was demonstrated by
\citet{FP07,fromang-etal-2007} in magneto-rotational turbulence.

As in \citet{oishi-maclow-2006}, we fail to detect a significant
change of regime in the spectra at the expected AD length scale
$\ell_a$, but in our simulations it happens at a greater length
scale. This length scale $\ell_a$ is predicted from the balance
between the moduli of the Fourier coefficients for the inertial
e.m.f. and the AD e.m.f.. We would underestimate $\ell_a$ if AD was
more coherent in time than the advection of the field, and so we
conjecture that AD terms have a greater coherence time than the
inertial terms.

In our simulations, we observe that AD shuts off the Lorentz force at
small scales: this shifts the peak of the AD heating power spectrum to
larger scales. The position of that peak defines a scale which seems
to match the characteristic thickness of the sheets where AD heating
is strong. This scale $\ell_a^*$ might be revealed by the
characteristic chemistry of AD heating where neutral-ion endothermic
reactions are favored (e.g.: CH$^+$ or SH$^+$ formation, see
\citealp{godard-2008}). Table \ref{table:ISM} sums up the
characteristics of various environments of the ISM. We identify the
integral length scale and the r.m.s. velocity in our simulations to
their corresponding physical values in each considered ISM components
to apply our results.  Although we find that $Re_a^{-1}$ in the ISM
varies from $10^{-3}$ to $10^{-2}$, we use $\ell_a^*=0.6$ as measured
in our AD simulations with $Re_a^{-1}=0.01$ to estimate the scale
$l_0\ell_a^*$ of the AD diffusion heating. The AD dissipation heating
is always about a fourth of the typical scale whereas $l_0.\ell_a$ can
be much smaller.

\begin{table*}
\caption{Characteristics of various components of the ISM. Dimensions
  are recovered from our simulations by assuming $L \simeq 2.5$,
  $\ell_a^*\simeq 0.6$ and $ u_0\simeq \sqrt{3}.U_a$ where $U_a$ is
  the line-of-sight r.m.s. Alfv\'en velocity. The quantity $\gamma
  \rho_i$ is computed by assuming the ions are essentially C$^+$ ions
  with a number density $n_i=10^{-4}n_\mathrm{H}$.
  \label{table:ISM}}
\begin{tabular}{ccccc}
  \hline\\
  &&CNM & molecular clouds & low-mass dense cores\\
  \hline\\
  Density & $n_\mathrm{H} ($cm$^{-3})$ & 30  & 200 & 10$^4$ \\
  Length scale & $L.l_0$ (pc)       & 10  & 3   & 0.1 \\
  r.m.s. velocity & $U.u_0/\sqrt{3}$ (km/s)     & 3.5 & 1   & 0.1 \\
  Alfv\'en velocity & $U_a=u_0/\sqrt{3} $ (km/s)     & 3.4 & 2   & 1 \\
  AD Reynolds number & $Re_a^{-1}= 1/\gamma\rho_i/(t_0)$ & 1.2~10$^{-2}$ & 3.6~10$^{-3}$ & 1.1~10$^{-3}$ \\
  AD heating length & $\ell_a^*.l_0$ (pc) & 2.4 & 0.72 & 0.024 \\
  AD dynamical length & $\ell_a.l_0=2\pi l_0Re_a^{-1}.U_a^2/U^2$ (pc) & 0.28 & 0.10 & 0.026 \\
  \hline\\
\end{tabular}
\end{table*}

The qualitative picture of the dissipation field suggests that it is
dominated by intermittent sheet-like structures which alternate with
large voids of low dissipation. The sheets of various dissipative
nature (Ohmic, viscous or AD) appear to be clearly separated, except
for the highest dissipation rates, where viscosity fades out and Ohmic
and AD heating blend. AD heating sheets are often seen to sandwich
much thinner regions of strong Ohmic or viscous dissipation, as in the
simple case studied by \citet{brandenburg-zweibel-1994}. The high
degree of intermittency is confirmed by the computation of the
structure function exponents for the velocity and the magnetic field,
as well as by the pdf of the total dissipation, which exhibits a
log-normal core and a strong power-law tail for high values of the
dissipation. We compared the statistics of the structures of strong
dissipation with those of UR10 and we obtain only marginal agreement,
probably because of the systematics linked with the definition of the
inertial and dissipative ranges. The statistical analysis of
structures of high total dissipation reveals the highly intermittent
nature of the dissipation field, as more than 30\% of the dissipation
takes place in less than 3\% of the volume. No significant difference
in the scaling laws between pure MHD and AD MHD was found, but the
slope of the power-law pdfs is affected in the dissipative range, with
a statistical preference towards more fragmented structures.

In future work, we intend to make progress towards a more realistic
picture of the ISM, relaxing the incompressible hypothesis, with the
final aim of including realistic cooling. We also hope our statistical
results will provide new ways to approach the observed characteristics
of the ISM, intermediate between direct post-processing of 3D
numerical simulations and the building of line of sights with
elementary models such as shocks or vortices \citep{godard-2008}.

\section*{Acknowledgments}
We thank V.M. Uritsky for sharing with us his experience with the
structures extraction algorithm. We would also like to thank the MesoPSL team of the Observatoire de Paris for providing support for the numerical simulations.

\bibliographystyle{plainnat} \bibliography{momferratos-et-al}

\end{document}